\documentclass[12pt,preprint]{aastex}
\usepackage{lineno}

\begin{document}

\title{Jet Radiation Properties of 4C +49.22: from the Core to Large-Scale Knots}
\author{ Jin Zhang\altaffilmark{1}, Hai-Ming Zhang\altaffilmark{2}, Su Yao \altaffilmark{3,1}, Sheng-Chu Guo\altaffilmark{2}, Rui-Jing Lu\altaffilmark{2}, En-Wei Liang\altaffilmark{2}}

\altaffiltext{1}{Key Laboratory of Space Astronomy and Technology, National Astronomical Observatories, Chinese Academy of Sciences, Beijing 100012, China; jinzhang@bao.ac.cn}
\altaffiltext{2}{Guangxi Key Laboratory for Relativistic Astrophysics, Department of Physics, Guangxi University, Nanning 530004, China}
\altaffiltext{3}{Kavli Institute for Astronomy and Astrophysics, Peking University, Beijing 100871, China; KIAA-CAS Fellow }

\begin{abstract}
4C +49.22 is a $\gamma$-ray flat spectrum radio quasar with a bright and knotty jet. We investigate the properties of the core and large-scale knots by using their spectral energy distributions (SEDs). Analyzing its \emph{Fermi}/LAT data in the past 8 years, a long-term steady $\gamma$-ray emission component is found besides bright outbursts. For the core region, the $\gamma$-ray emission together with the simultaneous emission in the low-energy bands at different epochs is explained with the single-zone leptonic model. The derived magnetization parameters and radiation efficiencies of the radio-core jet decrease as $\gamma$-ray flux decays, likely indicating that a large part of the magnetic energy is converted to the kinetic energy of particles in pc-scale. For the large-scale knots, their radio--optical--X-ray SEDs can be reproduced with the leptonic model by considering the inverse Compton scattering of cosmic microwave background photons. The sum of the predicted $\gamma$-ray fluxes of these knots is comparable to that observed with LAT at $\sim10^{24}$ Hz of the steady $\gamma$-ray component, indicating that the steady $\gamma$-ray emission may be partially contributed by these large-scale knots. This may conceal the flux variations of the low-level $\gamma$-ray emission from the radio-core. The derived bulk Lorentz factors of the knots decrease along the distance to the core, illustrating as deceleration of jet in large-scale. The powers of the core and knots are roughly in the same order, but the jet changes from highly magnetized at the core region into particle-dominated at the large-scale knots.
\end{abstract}

\keywords{galaxies: active---galaxies: jets---radiation mechanisms: non-thermal---galaxies: individual: 4C +49.22}

%%%%%%%%%%%%%%%%%%%%%%%%%%%%%%%%%%%%%%%%%%%%%%%%%%%%%%%%%%%%%%%%
\section{Introduction}           %% first-level sections will be auto-capitalized
\label{sect:intro}

The substructures of large-scale jets in radio-loud active galactic nuclei (AGNs), i.e., knots, hotspots, and lobes, have been resolved at the radio, optical and X-ray bands (see Harris \& Krawczynski 2006 for a review). This presents an opportunity to reveal the jet properties from the radio-core to the large-scale knots. This is helpful for revealing jet formation and propagation, composition, particle acceleration and radiation mechanisms, etc. (e.g., Zargaryan et al. 2017). The very long baseline interferometry (VLBI) observations in multi-epoch measurements of sub-parsec scale jets suggest that AGN jets start out highly relativistic with a Lorentz factor of tens (Jorstad et al. 2005; Lister et al. 2016), and it was proposed that the jets are still mildly relativistic at the kpc-scale (e.g., Arshakian \& Longair 2004; Mullin \& Hardcastle 2009). Convincing evidence for jet deceleration and transverse motions in M87 is presented by measuring its kpc-scale proper motions with Hubble Space Telescope (HST) and the pc-scale proper motions with VLBI, and the apparent velocity that is still superluminal in its outer jet (Meyer et al. 2017). The broadband spectral energy distributions (SEDs) of both the radio-core and large-scale jet radiations for radio-loud AGNs shows that they are non-thermal emission origin and show a bimodal feature (e.g., Sikora et al. 1994; Ghisellini et al. 2009; Zhang et al. 2010, 2013, 2014, 2018). The high energy radiation beyond the X-ray band of the core region should be dominated by the synchrotron-self-compton scattering (SSC, Sikora et al. 1994; Ghisellini et al. 2009; Zhang et al. 2013) process and/or the inverse scattering the photons of the broad-line region (IC/BLR, Ghisellini et al. 2009; Zhang et al. 2014, 2015) or torus (IC/torus, Sikora et al. 2009; Kang et al. 2014). The high energy (the X-ray--$\gamma$-ray bands) radiation mechanisms of large-scale jets are still debated (Harris \& Krawczynski 2006; Zhang et al. 2010; Meyer et al. 2015; Zargaryan et al. 2017).

4C +49.22 is a $\gamma$-ray flat spectrum radio quasar (FSRQ) at redshift $z=0.334$ (Burbidge 1968; Lynds \& Wills 1968). It has a one-side, knotty and wiggling jet, and its knots were resolved at the radio, optical and X-ray bands (Owen \& Puschell 1984; Akujor \& Garrington 1991; Sambruna et al. 2004, 2006). This source was not detected by the previous $\gamma$-ray detectors, such as EGRET (Hartman et al. 1999) and AGILE (Pittori et al. 2009), but a bright $\gamma$-ray outburst was detected with \emph{Fermi}/LAT (Reyes et al. 2011; Cutini et al. 2014). The outburst was also simultaneously observed from the microwave to the X-ray bands with \emph{Planck} and \emph{Swift}. The $\gamma$-ray flux is highly variable and correlated with the emission in the low energy bands, indicating that the $\gamma$-ray outburst is from the compact core region (Cutini et al. 2014). In addition, a new component from the radio-core around the time of the $\gamma$-ray outburst was catched with the Very Long Baseline Array (VLBA, Cutini et al. 2014). This robustly suggests that the outburst is in the vicinity of the core region and is related to the activities of the central black hole. Interestingly, a steady $\gamma$-ray emission component was detected with the \emph{Fermi}/LAT during the past 8 operation years. It is unclear whether the steady $\gamma$-ray emission component is attributed to the radio-core or the knots of this source. So far, the $\gamma$-ray emission outside the radio-core was only convincingly detected by the \emph{Fermi}/LAT for the radio lobes of Cen A (Abdo et al. 2010) and Fornax A (McKinley et al. 2015; Ackermann et al. 2016). If the steady $\gamma$-ray emission of 4C + 49.22 is from the knots, it wound be added as a valuable source with $\gamma$-ray emission at the large-scale jet structure.

This paper dedicates to study the emission mechanisms of the radio-core and knots for 4C +49.22 for revealing the jet properties from the radio-core to large-scale knots. We analyzed the observational data of \emph{Fermi}/LAT for 4C +49.22 in the past 8 years, and the derived $\gamma$-ray light curve is presented in Section 2. We model the broadband SEDs of the core region at different epochs with a single-zone leptonic model in the IC/BLR scenario (Section 3.1). We also model the SEDs in the radio--optical--X-ray band for the knots with the leptonic model in the IC/CMB scenario and compare the $\gamma$-ray flux predicted by the model to the steady $\gamma$-ray component of the LAT observation (Section 3.2). The jet properties from the core region to the large-scale knots are presented in Section 4. Discussion and a summary are given in Section 5 and Section 6, respectively. Throughout, $H_0=71$ km s$^{-1}$ Mpc$^{-1}$, $\Omega_{\rm m}=0.27$, and $\Omega_{\Lambda}=0.73$ are adopted.

\section{\emph{Fermi}/LAT Data Reduction and Description}

We downloaded the \emph{Fermi}/LAT data of 4C + 49.22 covering from 2008 August 6 (Modified Julian Day, MJD 54684) to 2018 January 24 (MJD 58142) from the Fermi data archive (Pass 8 data). The reduction and analysis of \emph{Fermi}/LAT data were performed with the standard analysis tool \emph{gtlike/pyLikelihood}, which is part of the Fermi Science Tool software package (ver. v10r0p5). The P8R2-SOURCE-V6 set of instrument response functions (IRFs) was used. Photons with energies from 0.1 to 100 GeV are taken into account for our analysis. The significance of the $\gamma$-ray signal from the source is evaluated with the maximum-likelihood test statistic (TS). The events are selected from the region of interest (ROI) with radius of 10$^{\circ}$, centered at the position of 4C +49.22. All point sources in the third \emph{Fermi}/LAT source catalog located in the ROI and an additional surrounding 10$^{\circ}$ wide annulus were modeled in the fits. In the model file, the spectral parameters for sources lying within the ROI were kept free and for sources lying within the annulus were fixed. The isotropic background, including the sum of residual instrumental background and extragalactic diffuse $\gamma$-ray background, was fitted with a model derived from the isotropic background at high Galactic latitude, i.e., ``iso-P8R2-SOURCE-V6-v06.txt", and the Galactic diffuse GeV emission was modeled with ``gll-iem-v06.fits". In order to eliminate the contamination from the $\gamma$-ray-bright Earth limb, the events with zenith angles $>100^{\circ}$ were excluded. The spectral analysis in the energy range of 0.1--100 GeV was performed by using the \textit{unbinned likelihood analysis}. A power law (PL) function, i.e., $dN(E)/dE = N_{\rm p}(E/E_{\rm p})^{-\Gamma_{\gamma}}$, is used to fit the spectrum accumulated in each time-bin.

\begin{figure*}
\includegraphics[angle=0,scale=0.45]{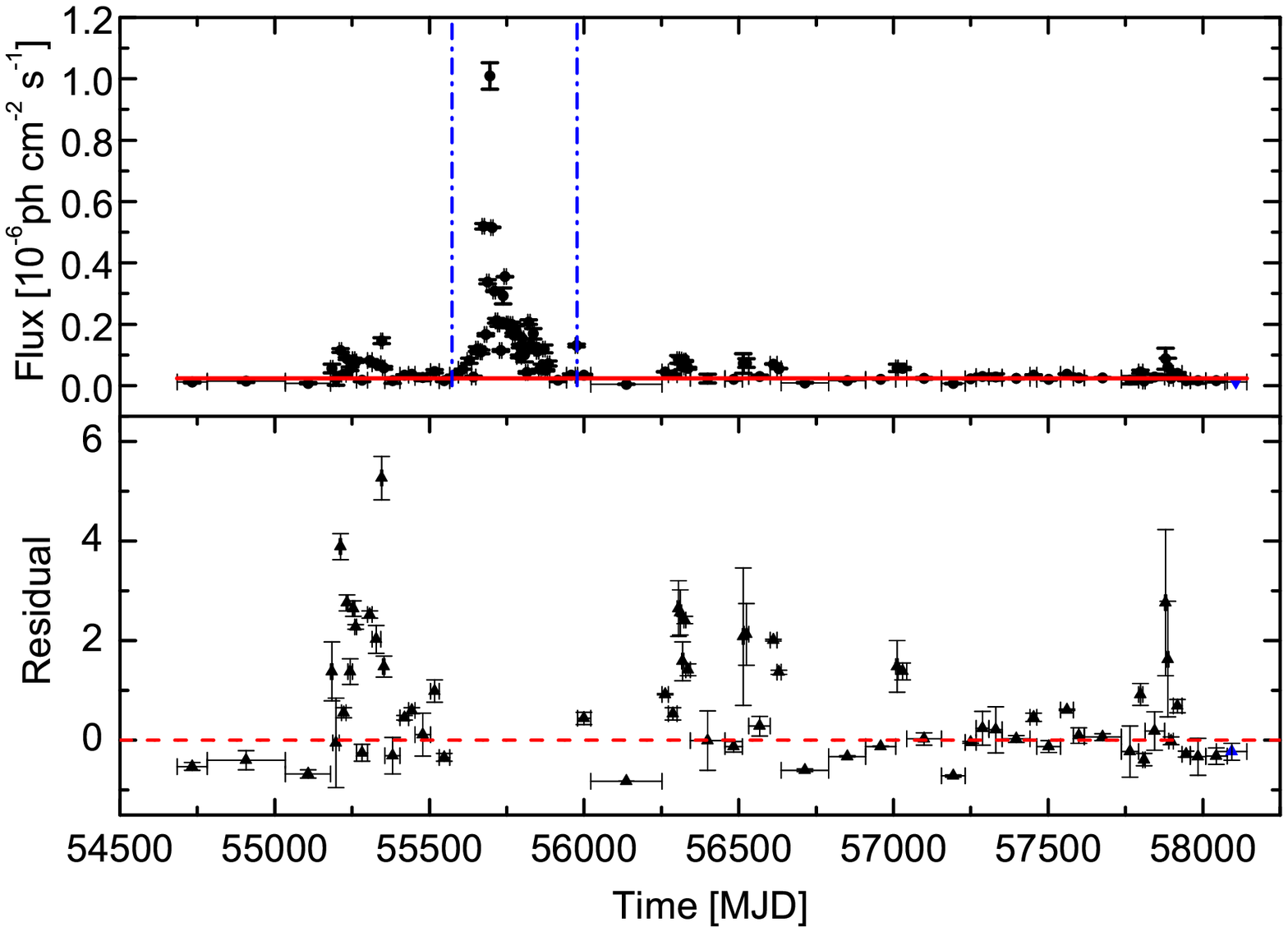}
\caption{Light curve observed by the \emph{Fermi}/LAT with a constant-significance of TS$\geq9$ in each adaptive time-bin (top panel). The red horizontal solid line is the 6-year average flux of $F_{\gamma,\rm c}= (2.3\pm0.1)\times10^{-8}$ photons cm$^{-2}$ s$^{-1}$ for the steady emission, i.e., from MJD 55978 to MJD 58142. The outburst (MJD 55573-55978) is marked by the blue vertical dotted lines. The fitting residuals of the steady emission by the 6-year average flux are given in the bottom panel. The blue symbols for the last time-bin indicates its TS$<9$. }\label{LC}
\end{figure*}

We do not use an even time-bin to make the \emph{Fermi}/LAT light curve, but adopt an adaptive-binning method to generate the light curve (Lott et al. 2012). The criterion for the time size selection is taken as TS$\geq$ 9, where TS$=9$ corresponds to $\sim3\sigma$ detection ( Mattox et al. 1996). The derived light curve is shown in Figure \ref{LC}. One can observe that the light curve is roughly composed of a long-term, steady $\gamma$-ray emission component and a bright outburst lasting about one year (from MJD 55573 to MJD 55978), which was reported by Hays \& Donato (2011). The flux of the steady $\gamma$-rays are low-level, showing as almost a constant flux. With the six-year LAT data post the outburst (MJD 55978-58142), we get an average flux as $F_{\gamma,\rm c}= (2.3\pm0.1)\times10^{-8}$ photons cm$^{-2}$ s$^{-1}$. The flux variation along this flux level with the residual defining as residual$\equiv (F_\gamma-F_{\gamma, \rm c})/F_{\gamma, \rm c}$ during the 8-year LAT observations is also shown in the low panel of Figure \ref{LC}, where the residuals of the outburst data are excluded for better illustrating the flux variation of the steady $\gamma$-ray component. One can observe that this steady $\gamma$-ray component has a temporal coverage from prior the outburst to late epoch after the outburst, and flux variations with a residual being greater than 3 are also seen in some time-bins.

\section{SED Modeling and Results for the Core and Knots }

As reported by Cutini et al. (2014), the $\gamma$-ray outburst should occur in the core region. This is also confirmed by observations of a new component from the radio-core around the time of the $\gamma$-ray outburst with the VLBA. The simultaneously observed SEDs of the core region during the flare (MJD 55695--55697) and at post-flare (MJD 55698--55706) and the archival low state data of the core region are taken from Cutini et al. (2014), as shown in Figure \ref{core}. In large-scale, eight knots of the 4C + 49.22 jet are resolved at the radio and X-ray bands, but no optical data are available for knots F, G, H (Sambruna et al. 2006). The SED data of the eight knots are taken from Sambruna et al. (2006). The average spectra in the \emph{Fermi}/LAT band during the steady $\gamma$-ray emission (MJD 55978--58142) and during the second \emph{Fermi}/LAT source catalogue (the same data in Figure \ref{core}(c), approximatively being the average spectrum before the outburst) are also shown in the SEDs of these knots. Although we cannot confirm that they are indeed from the knots, they place constraint on our SED fits.

The single-zone synchrotron+IC radiation models are used to reproduce the broadband SEDs of the core region and knots. The radiation region is assumed to be a sphere with radius of $R$ and magnetic field strength of $B$. The bulk Lorentz factor of the emission region is $\Gamma$, and the beaming factor should be $\delta=1/\Gamma(1-\beta \cos\theta)$, where $\theta$ is the viewing angle. The radiation electron distribution is taken as a broken power-law (Ghisellini et al. 2009; Zhang et al. 2014, 2015; Chen et al. 2012), and this distribution is characterized by an electron density parameter ($N_0$), a break energy ($\gamma_{\rm b}$) , and two slope indices ($p_1$ and $p_2$) below and above the break energy in the energy range of $\gamma_{\rm e}\in[\gamma_{\rm min},\gamma_{\rm max}]$. The Klein-Nashina (KN) effect and the absorption of high energy $\gamma$-ray photons by extragalactic background light (Franceschini et al. 2008) are also taken into account in our model calculations.

\subsection{The Core Region}

\begin{figure*}
\includegraphics[angle=0,scale=0.45]{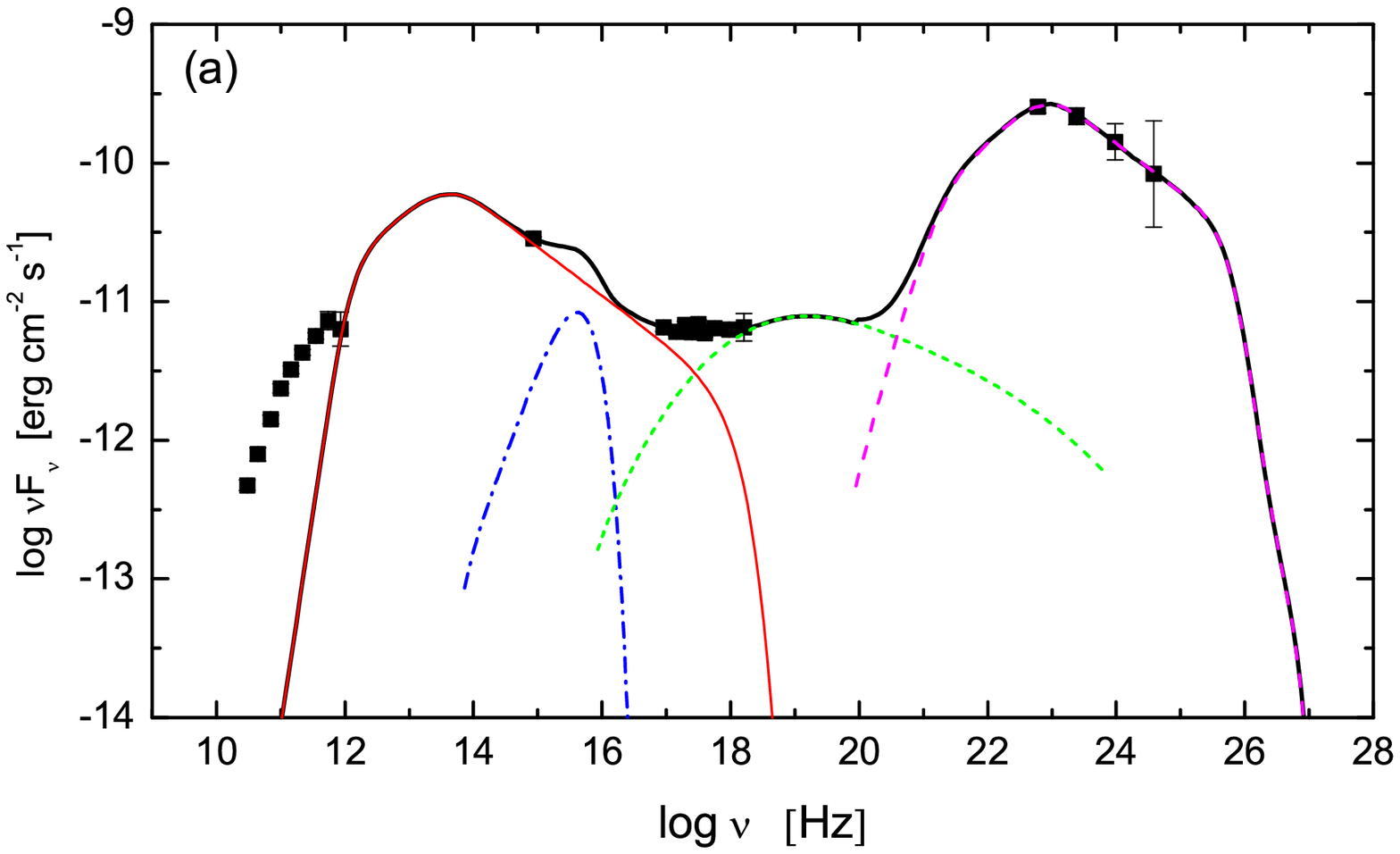}
\includegraphics[angle=0,scale=0.45]{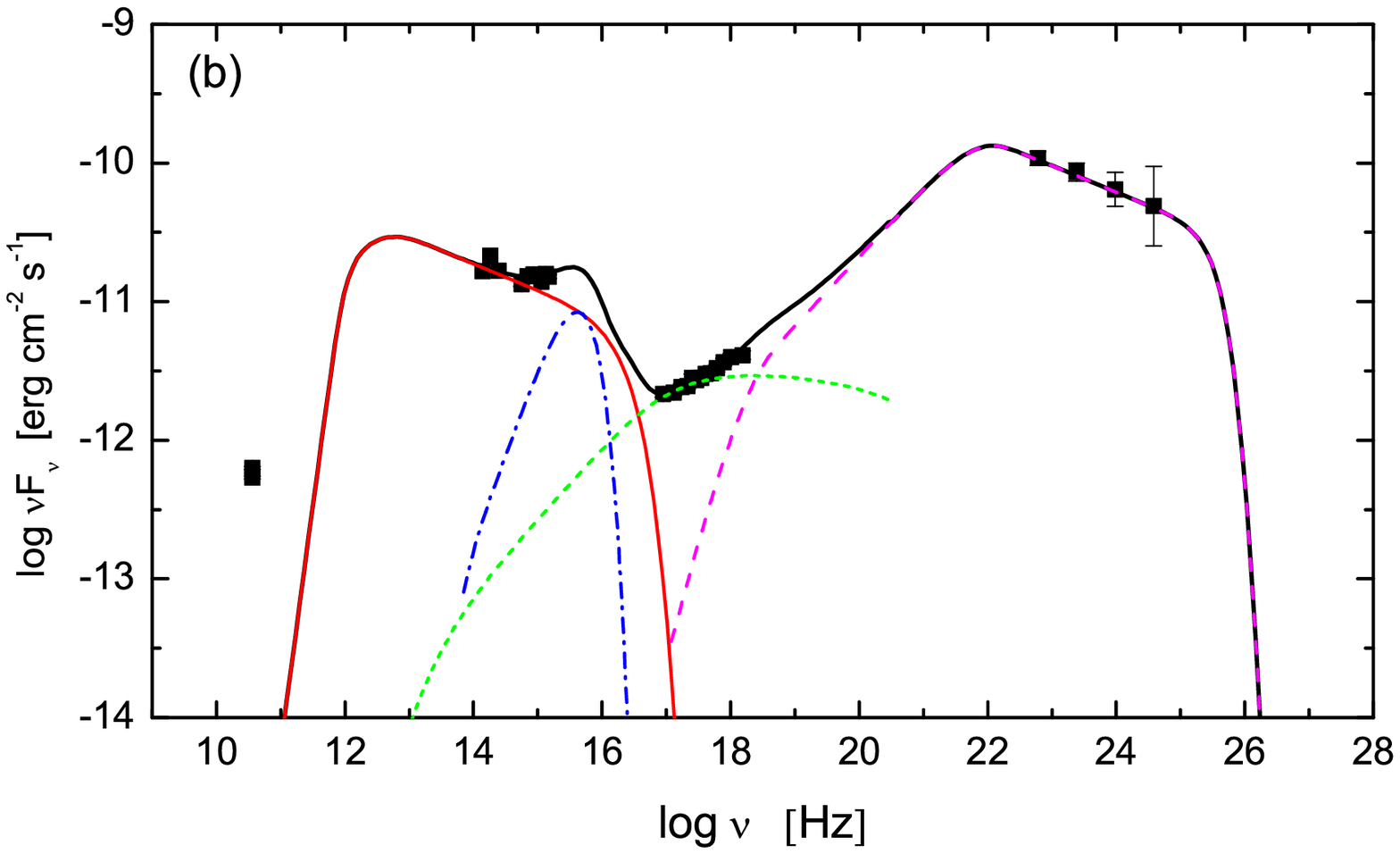}
\includegraphics[angle=0,scale=0.45]{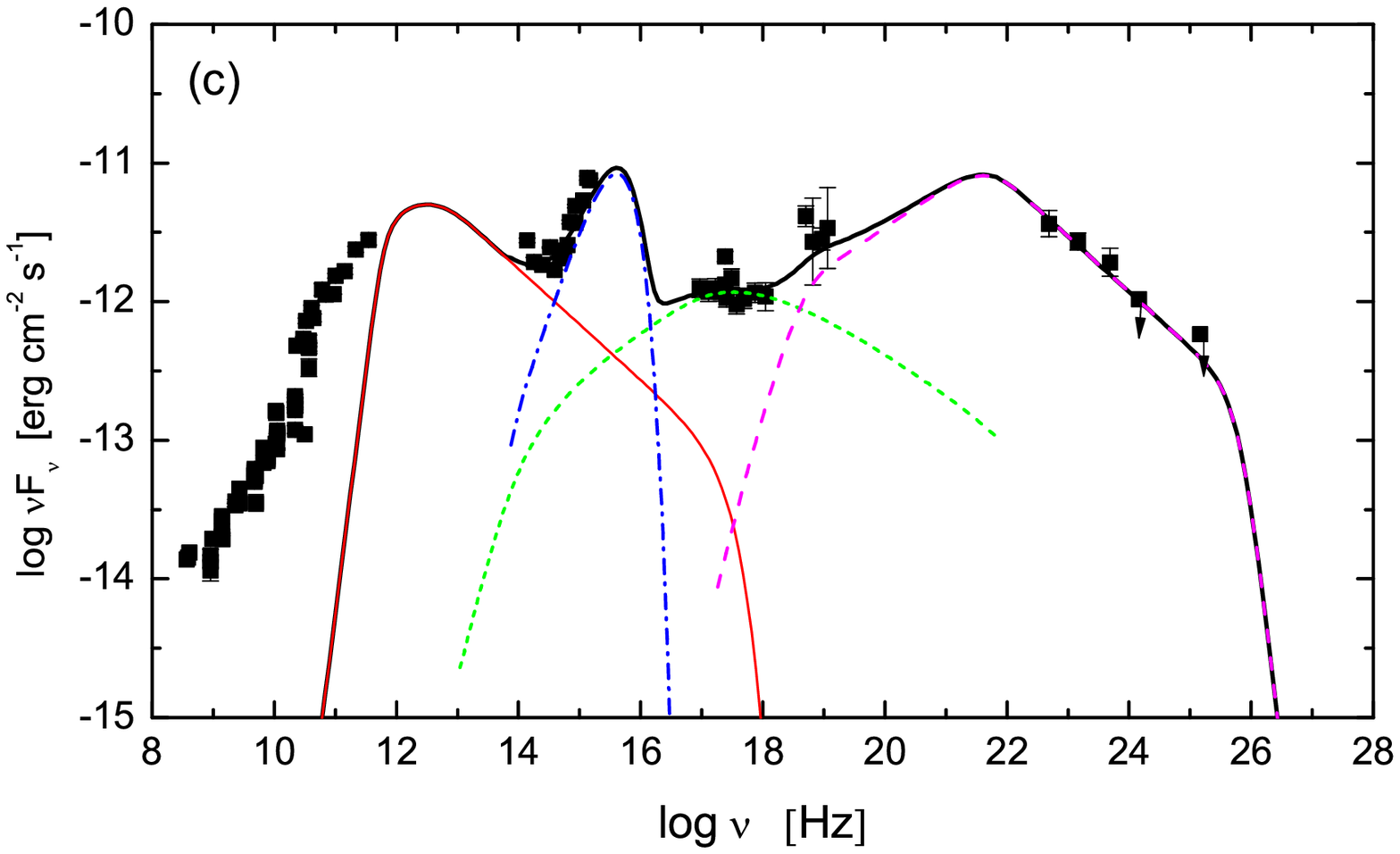}
\caption{Observed SEDs and model fitting lines for core region at different epochs, i.e., during flare (MJD 55695-55697, panel (a)), post flare (MJD 55698-55706, panel (b)), and archival low state (panel (c)). The data are taken from Cutini et al. (2014). The black thick solid lines are the sum of emission from synchrotron radiation (red thin solid lines), accretion disk (blue dash-dotted lines), SSC process (green shot-dashed lines), and IC/BLR process (magenta dashed lines).}\label{core}
\end{figure*}

As illustrated in Figure \ref{core}, the blue bump of the thermal emission from the accretion disk is prominent when the source is in a low state of $\gamma$-ray radiation. Following our previous work (Sun et al. 2015; Zhang et al. 2015), the standard accretion disk spectrum (Davis \& Laor 2011) is used to explain this thermal emission. The fitting parameters include the inside ($R_{\rm in}$) and outside radii ($R_{\rm out}$) of the accretion disk, black hole mass ($M_{\rm BH}$), Eddington ratio, and inclination to the line of sight $i$. The inner radiative edge of the accretion disk may be at the marginally stable orbit radius and outside the Schwarzschild radius ($R_{\rm s}$; e.g., Krolik \& Hawley 2002). We take $R_{\rm in}=R_{\rm s}$, $R_{\rm out}=500R_{\rm s}$, $\cos i = 1$, and $M_{\rm BH}=4.0\times10^8M_{\odot}$ (Shields et al. 2003) in this analysis. We vary the Eddington ratio to model the accretion disk emission in the ultraviolet band when the source is in a low state of $\gamma$-ray radiation, as shown in Figure \ref{core}(c). However, the thermal emission from the accretion disk is overwhelmed by the non-thermal emission of jet when the source is in a high state of $\gamma$-ray radiation. The variability timescale produced by the variations of the accretion rate is on timescale of years (e.g., Zhang et al. 2013; Smith et al. 2018). Therefore, the thermal emission component of the accretion disk is fixed during the SED fitting of the source in flare (Figure \ref{core}(a)) and post-flare state (Figure \ref{core}(b)).

We assume that the radiation region is inside the BLR, which is also consistent with the short timescale of variability at the GeV band. The SSC and IC/BLR processes are taken into account in our modeling. This model is widely used and can well represent the most observed SEDs of FSRQs (e.g., Sikora et al. 1994; Ghisellini et al. 2009; Chen et al. 2012; Zhang et al. 2014, 2015). The energy density of BLR at rest-frame is estimated by $U_{\rm BLR}=\frac{L_{\rm BLR}}{4\pi R^{2}_{\rm BLR}c}=0.046$ erg cm$^{-3}$, where $R_{\rm BLR}=1.26\times10^{17}$ cm (Decarli et al. 2008) is the radius of BLR and the BLR luminosity ($L_{\rm BLR}$) is estimated using the fluxes of emission lines reported in Table 3 and Equation (1) in Celotti et al. (1997). In the comoving frame, the energy density of BLR is boosted by a factor of $\Gamma^2$ and a corrected factor of 17/12 (Ghisellini \& Madau 1996) should be considered. Therefore, the energy density of the BLR photon fields in the comoving frame is  $U^{'}_{\rm BLR}=\frac{17}{12}\Gamma^2U_{\rm BLR}$. The spectrum of BLR can be approximated by a blackbody with a peak in the comoving frame at $2\times10^{15}\Gamma$ Hz (Ghisellini \& Tavecchio 2008).

The radiation region size of the core region is assumed to be a sphere with $R=\delta c\Delta t/(1+z)$, where $\Delta t$ is taken as 0.33 day (the rapid variability timescale in Cutini et al. 2014), 12 hr (the value for other FSRQs in Zhang et al. 2014, 2015), and 24 hr during SED fitting for flare, post-flare, and low states, respectively. We use the $\chi^{2}$ minimization technique to search the goodness of the SED fits, and the details of this technique please refer to Zhang et al. (2014, 2015). We first take $\delta=\Gamma$ in the model calculation for modeling the SED of the $\gamma$-ray flare. This means that the viewing angle is equal to the beaming angle of a jet, i.e., 1/$\Gamma$. We obtain $\delta=\Gamma=13.2$, yielding  a viewing angle of $\theta\sim4.3^{\circ}$, which satisfies the constraint of $\theta<11^{\circ}$ derived from the measured apparent superluminal speed of 9.9$c$ (Cutini et al. 2014). For consistency, we fix the viewing angle as $\theta\sim4.3^{\circ}$ during SED modeling for post-flare and low states.

The results of our SED fits are shown in Figure \ref{core} and the derived model parameters with 1$\sigma$ confidence level are reported in Table 1. The radio emission cannot be fitted due to the synchrotron self-absorption, and thus they may be from the larger radiation regions. The radiation at the X-ray band is mostly contributed by the SSC process while the $\gamma$-ray emission is represented by the IC/BLR process. It is found that the model fits appear appropriate to represent the observed SEDs of 4C +49.22 at different epochs.

As shown in Table 1, the derived values of B and $\Gamma$ decrease with the decrease of the $\gamma$-ray emission flux. The jet may be launched with strong magnetic fields, and then the magnetic energy converts into kinetic energy with the increase of bulk speed at pc-scale (see Boccardi et al. 2017 for a review; Chen 2018). Hence, the decrease of B and $\Gamma$ may indicate that the radiation region is away from the central engine following the decay of the $\gamma$-ray emission flux.

It is found that the peak frequencies of both synchrotron and EC bumps ($\nu_{\rm s}$ and $\nu_{\rm c}$) decrease following the decay of the $\gamma$-ray flux, i.e., from $4.63\times10^{13}$ Hz to $3.10\times10^{12}$ Hz for $\nu_{\rm s}$ and from $9.62\times10^{22}$ Hz to $4.17\times10^{21}$ Hz for $\nu_{\rm c}$. This tendency has been reported by Cutini et al. (2014), and the similar results have also been observed in other blazars (e.g., Zhang et al. 2013). As illustrated in Figure \ref{core}, the thermal emission of accretion disk is overwhelmed by the non-thermal emission of jet during the high $\gamma$-ray emission states. This phenomenon is also presented in other FSRQs, e.g., 3C 454.3 (Bonnoli et al. 2011). In addition, some sources are classified as BL Lacs but show some properties similar to FSRQs (Sbarufatti et al. 2005; Raiteri et al. 2007; Ghisellini et al. 2011; Giommi et al. 2012), which may be due to their intrinsically weak broad lines being overwhelmed by the beamed non-thermal continuum. Therefore, 4C +49.22 was suggested to display some features more typical of BL Lacs and support a smooth transition between the division of blazars into BL Lacs and FSRQs (Cutini et al. 2014).

\subsection{The Knots}

The radio emission of knots, including the optical radiation sometimes, should be of a synchrotron origin (e.g., Harris \& Krawczynski 2006). As suggested by Sambruna et al. (2006), the X-ray fluxes of some knots in 4C +49.22 are well above the extrapolation from the radio--to--optical spectra and the X-ray spectra are harden, indicating that an IC component is necessary to explain their X-ray emission. We therefore model the SEDs of the knots in the radio, optical and X-ray bands by considering both the synchrotron radiation and the IC process. As reported in Zhang et al. (2018), except for knot-IJ, if the X-ray emission is produced by the SSC process, the derived magnetic field strengths for other four knots are smaller than 1 $\mu$G, i.e., smaller than the magnetic field strength of interstellar medium. Hence the IC scattering of cosmic microwave background (IC/CMB) is suggested to produce the X-ray emission of knots (see also Sambruna et al. 2006; Tavecchio et al. 2000; Kataoka \& Stawarz 2005; Harris \& Krawczynski 2006; Zhang et al. 2009, 2010, 2018). The observed radio spectral indices are comparable to the X-ray slopes for the knots of 4C +49.22, which is also consistent with the model prediction if the same population of electrons contributes the radiations at the two energy bands by synchrotron and IC/CMB processes, respectively. Although the IC/CMB scenario requires highly relativistic outflows ($\Gamma\sim10$), implying that the jet does not suffer sever deceleration between pc and kpc scales, it is also favored by some observations; the high X-ray to radio luminosity ratio for an intermediate redshift ($z =2.5$) quasar B3 0727+490 appears consistent with the $\propto(1+z)^4$ amplification (Simionescu et al. 2016) and the X-ray--to--radio flux ratios of the high-redshift jets being marginally inconsistent with those from lower redshifts (McKeough et al 2016), which are expected from the IC/CMB model for the X-ray emission.

Although the contribution of SSC process is negligible comparing with IC/CMB component, we still take SSC process into account in our calculations. The single-zone synchrotron+SSC+IC/CMB model is used to fit the SEDs of knots. The CMB peak frequency at $z=0$ is $\nu_{\rm CMB}=1.6\times10^{11}$ Hz and the CMB energy density in the comoving frame is $U^{'}_{\rm CMB}=\frac{4}{3}\Gamma^2U_{\rm CMB}(1+z)^4$ (Dermer \& Schlickeiser 1994), where $U_{\rm CMB}=4.2\times10^{-13}$ erg cm$^{-3}$. For the SED fitting of these knots, we also take a viewing angle of $\theta\sim4.3^{\circ}$ and adjust the value of $\Gamma$ to get the value of $\delta$ during SED modeling, which is different from the assumption in Zhang et al. (2010, 2018).

The radius of radiation region is derived from the angular radius at the X-ray band and they are taken from Sambruna et al. (2006). However, only three knots (knot-B, knot-C, knot-E) have the observational data available, so we take the same value for knot-D and knot-C while the value of knot-E is used for the other four knots, as listed in Table 1. In order to constrain the model parameters, we assume that these knots satisfy the minimum energy condition, i.e., the energy densities of relativistic electrons and magnetic fields are in equipartition (see also Sambruna et al. 2006). Therefore, the magnetic field strength is estimated by $\frac{B^{2}}{8\pi}=U^{'}_{\rm e}$, where $U^{'}_{\rm e}$ is the energy density of relativistic electrons in the comoving frame.

\begin{figure*}
\includegraphics[angle=0,scale=0.28]{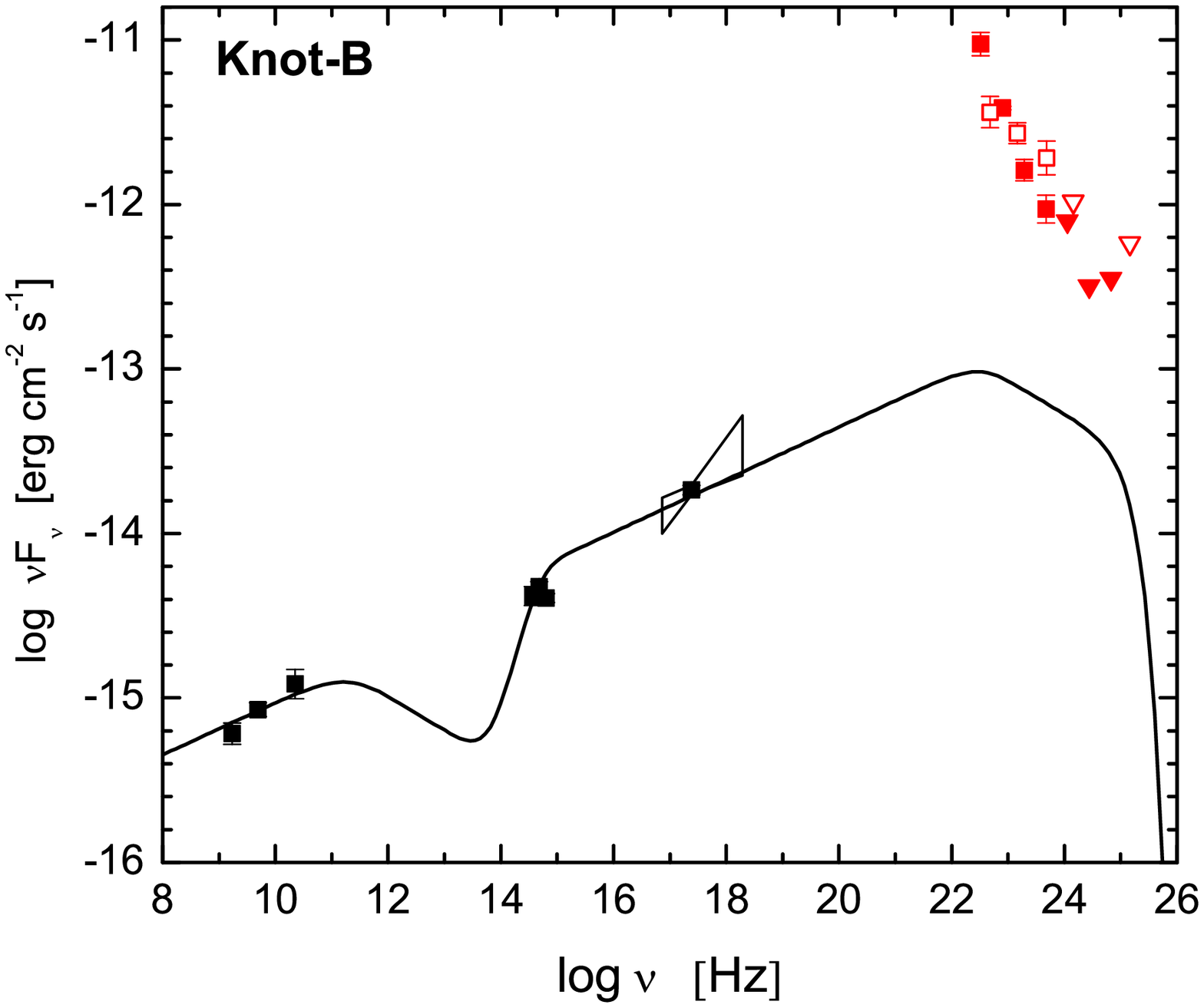}
\includegraphics[angle=0,scale=0.28]{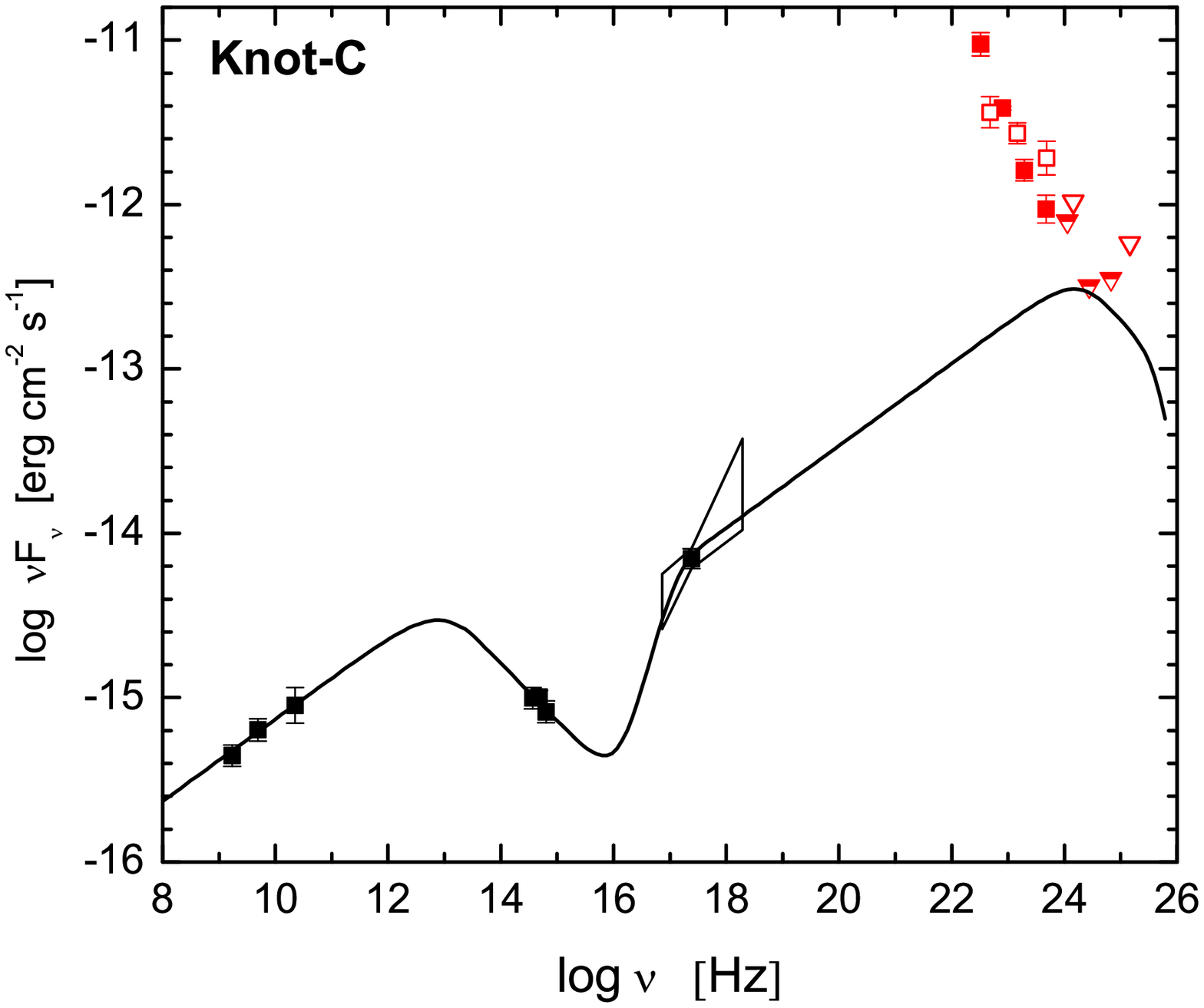}
\includegraphics[angle=0,scale=0.28]{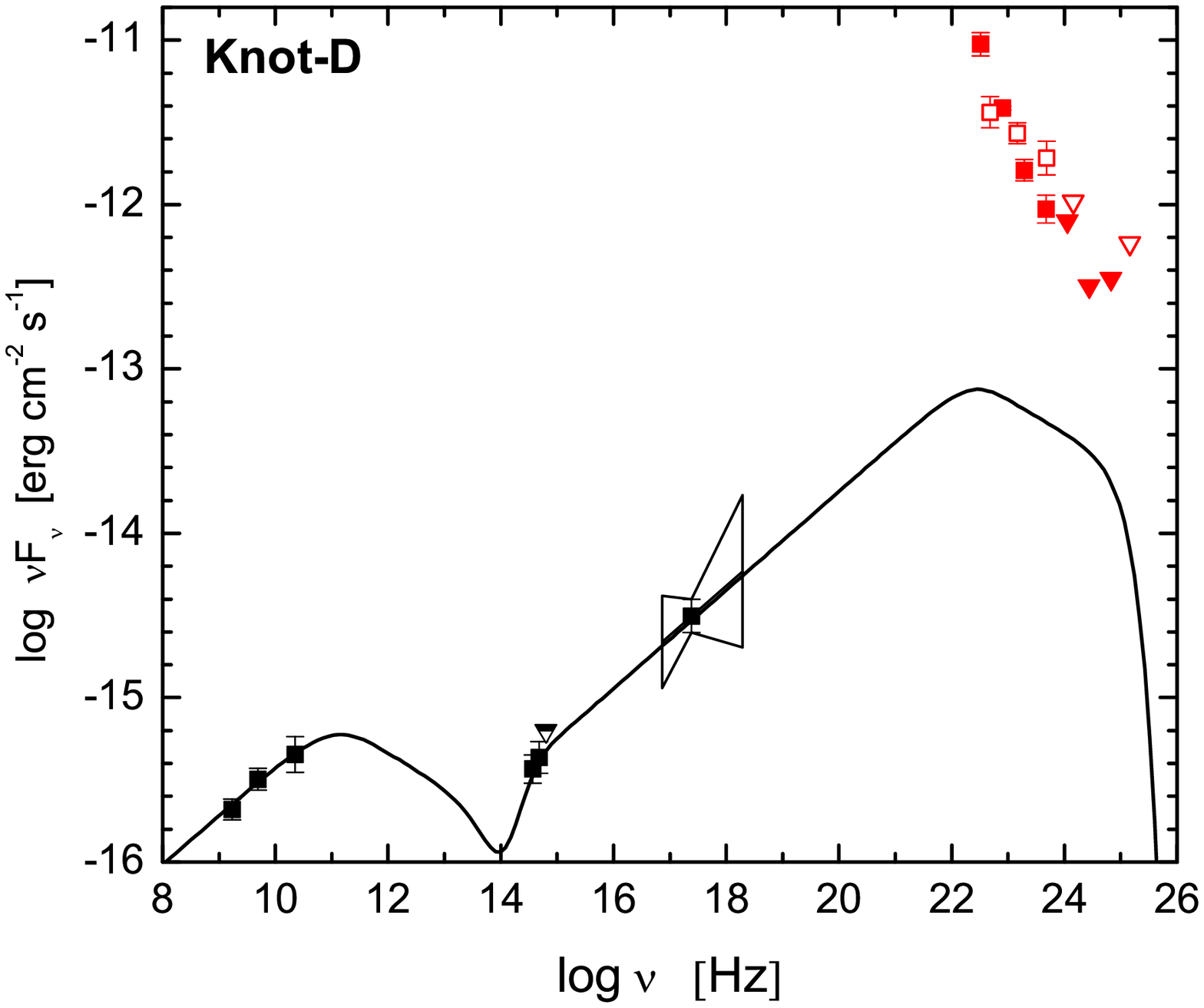}\\
\includegraphics[angle=0,scale=0.28]{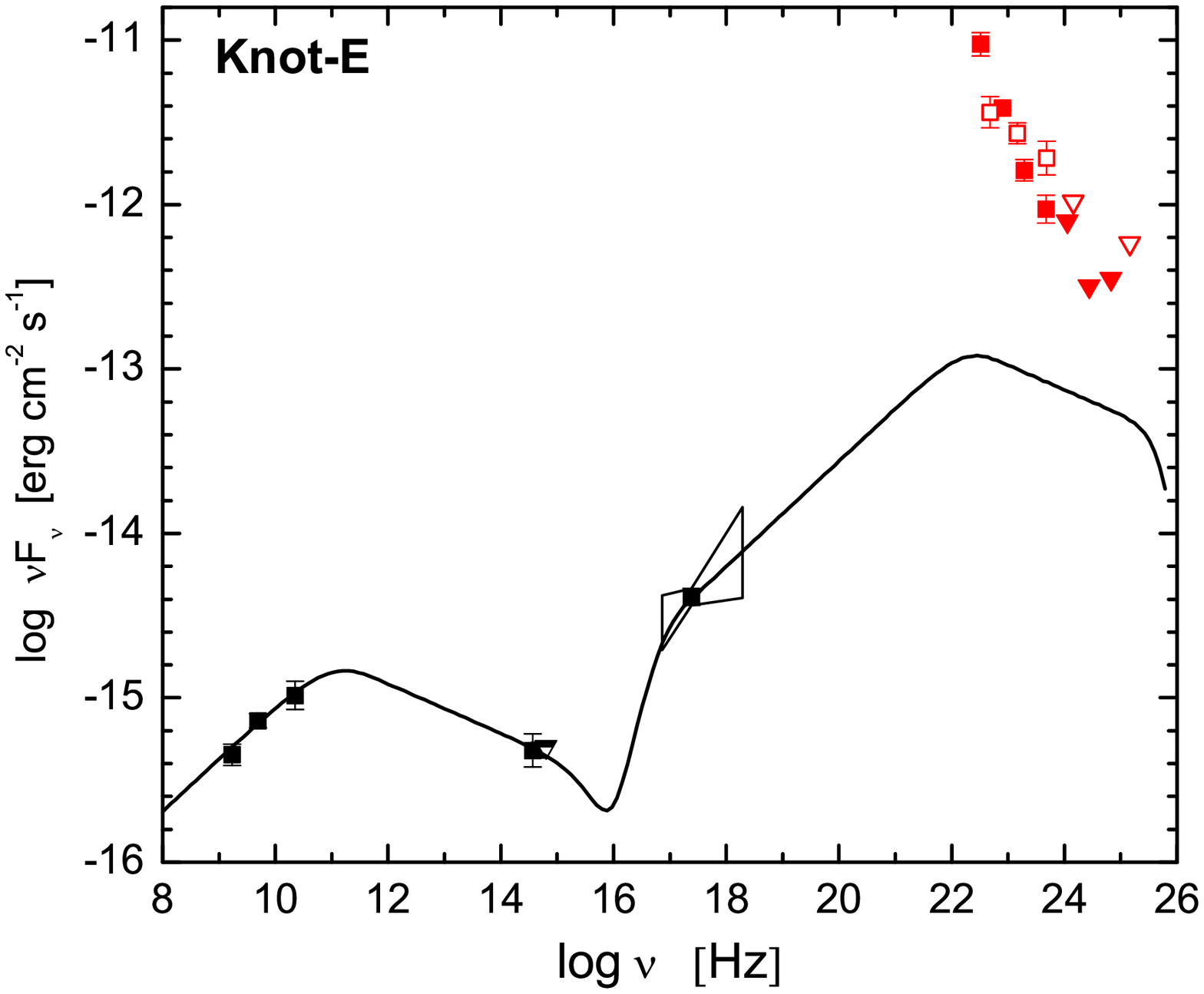}
\includegraphics[angle=0,scale=0.28]{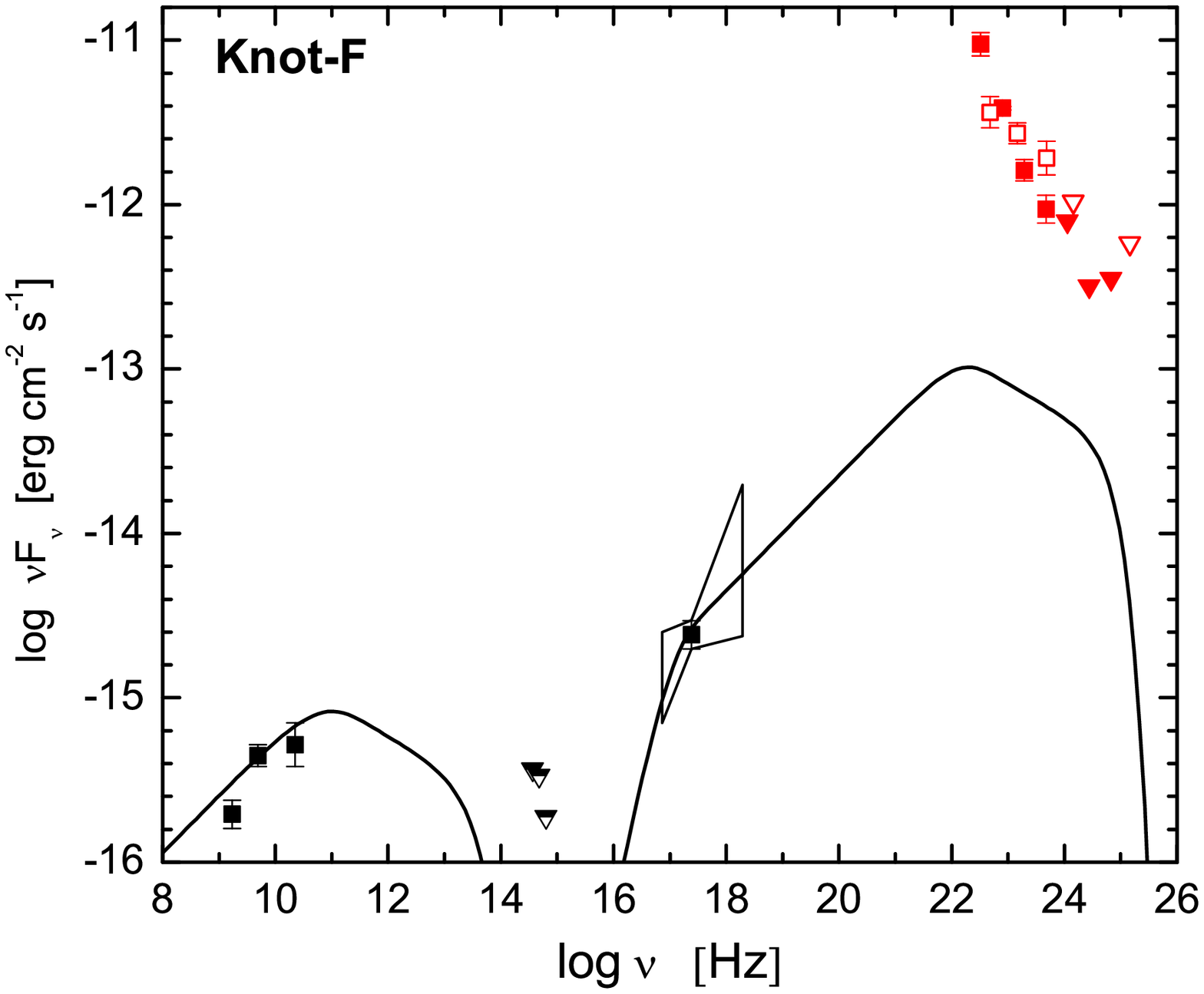}
\includegraphics[angle=0,scale=0.28]{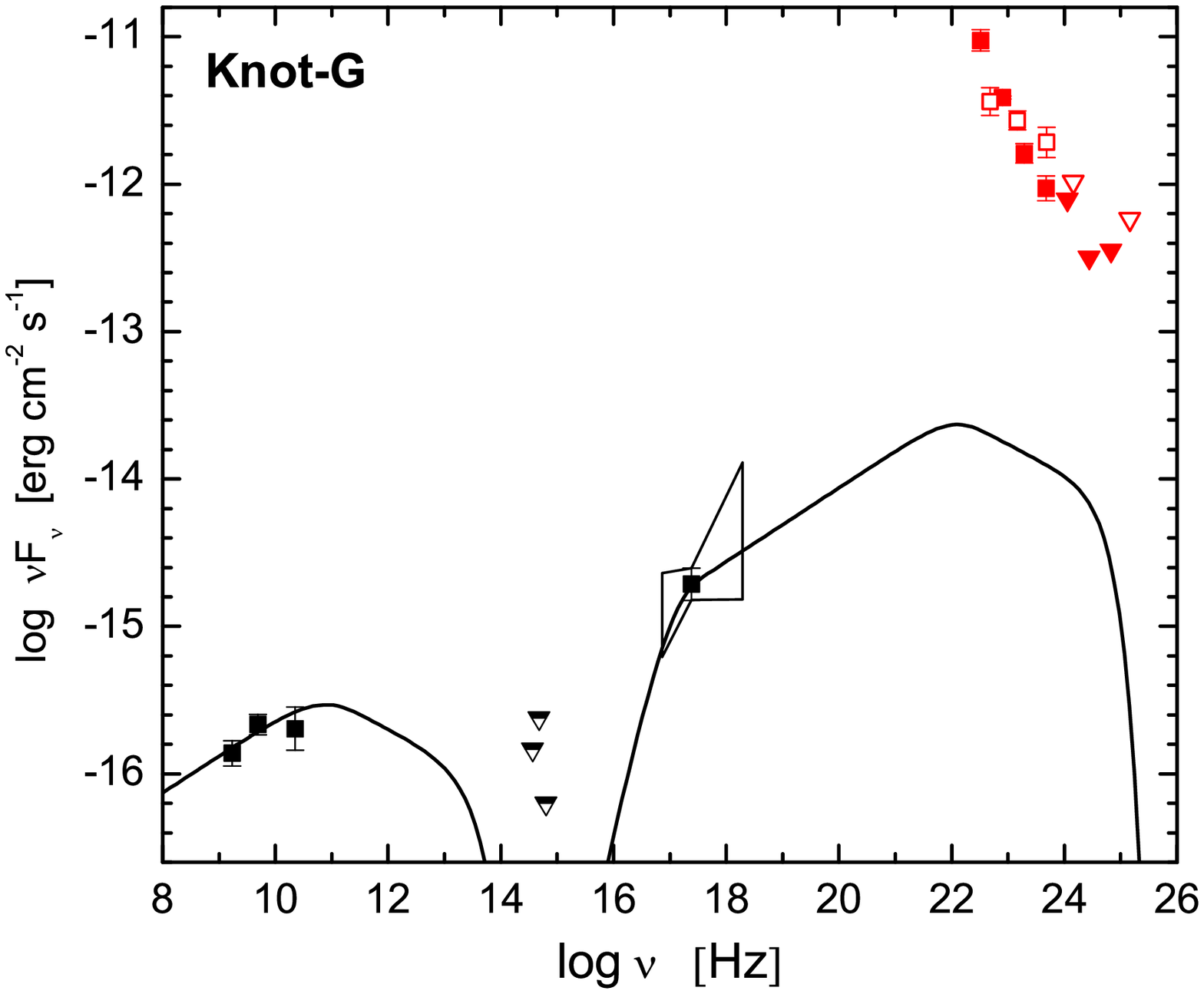}\\
\includegraphics[angle=0,scale=0.28]{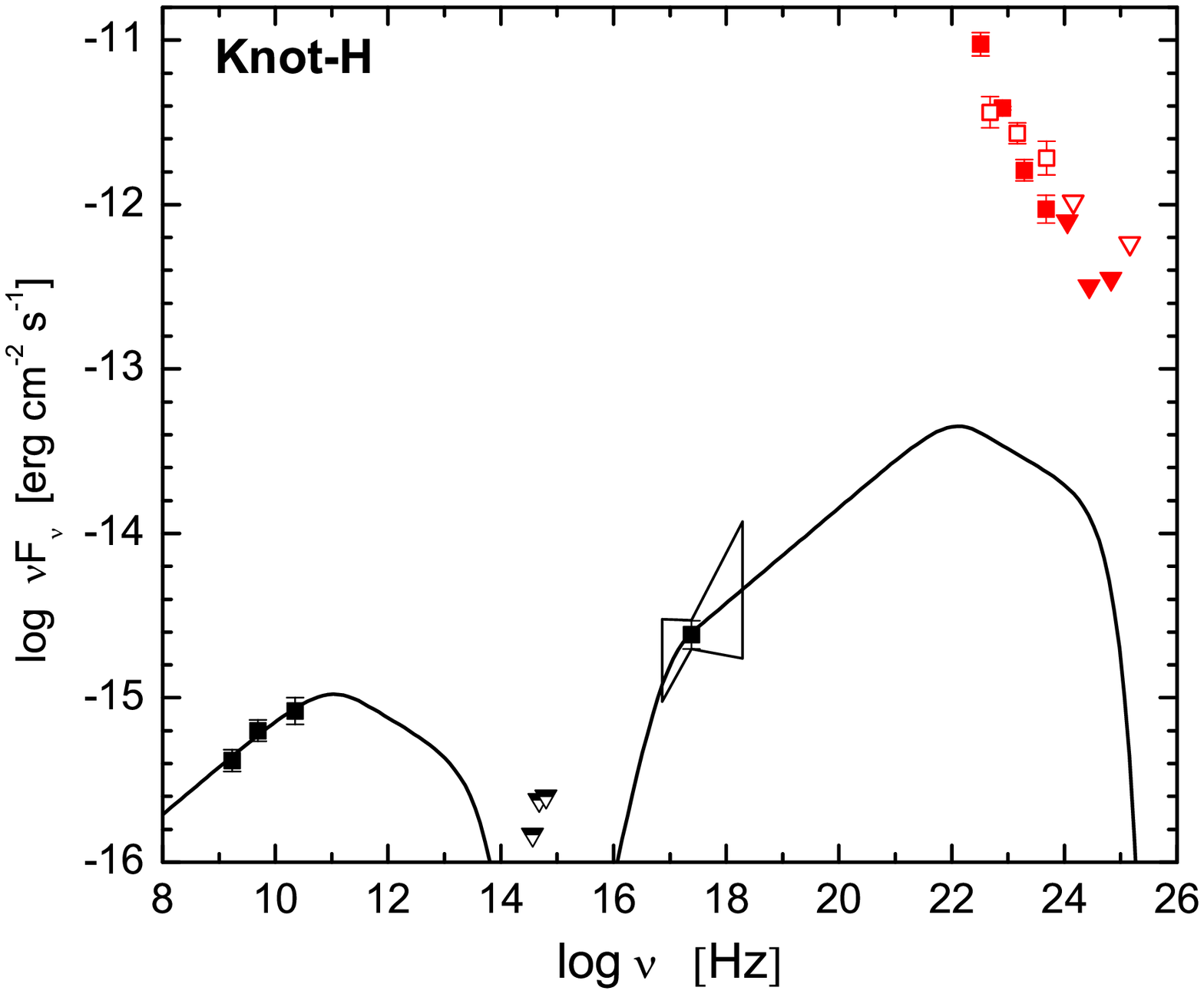}
\includegraphics[angle=0,scale=0.28]{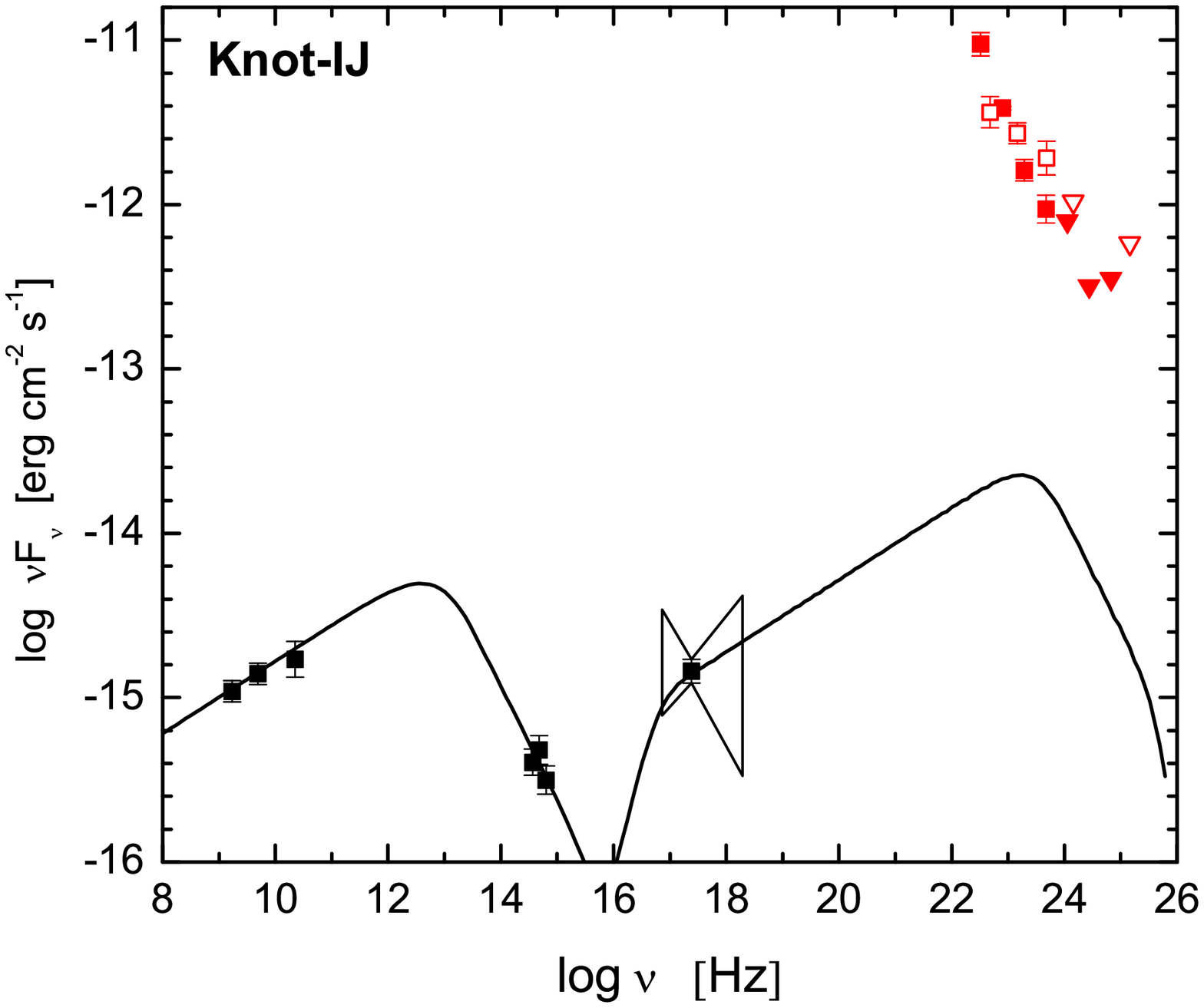}
\includegraphics[angle=0,scale=0.28]{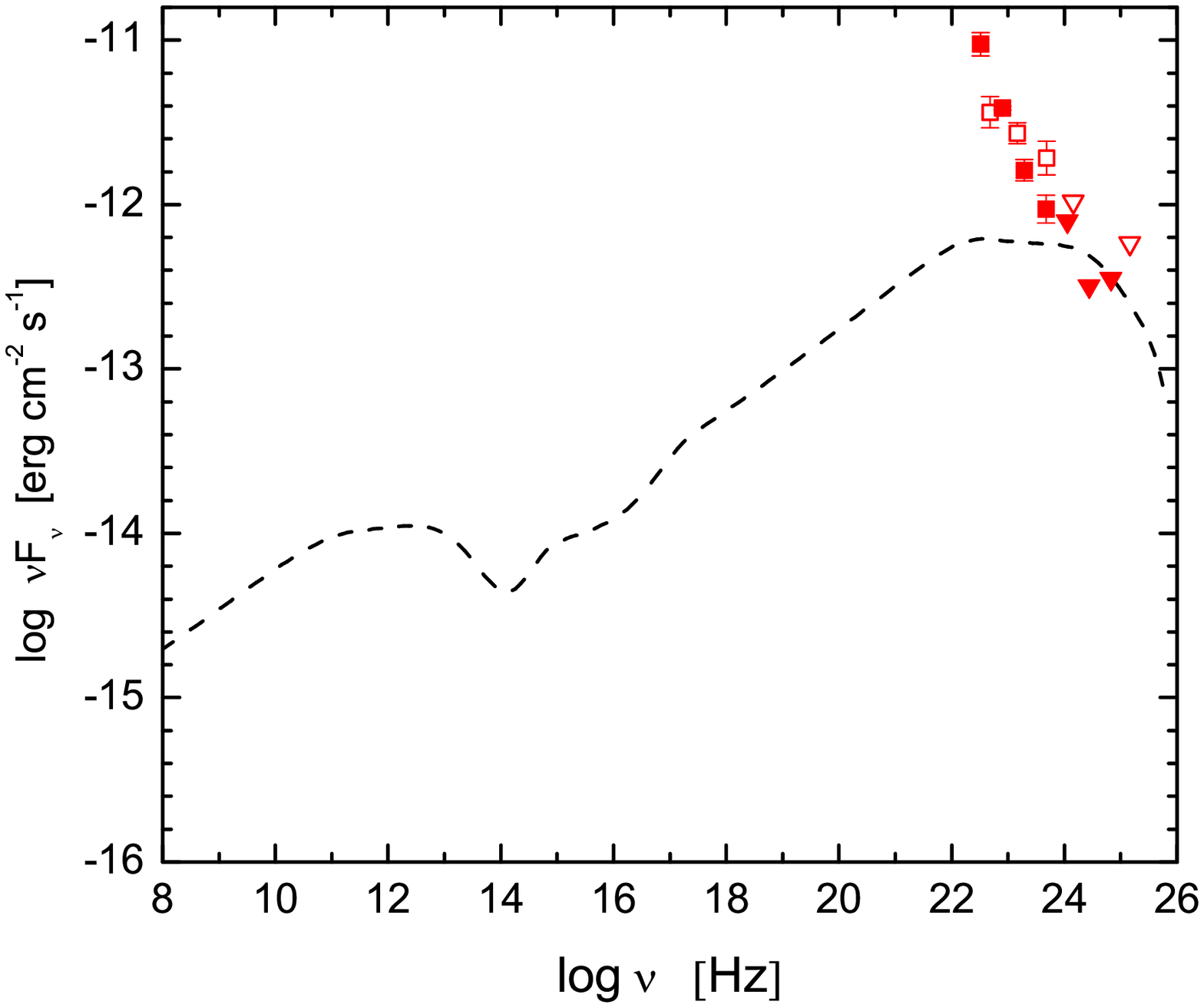}
\caption{Observed SEDs and model fitting lines for knots. The red symbols at the $\gamma$-ray band indicate the average spectra of the \emph{Fermi}/LAT observations during the steady state (MJD 55978--58142, solid symbols) and the second \emph{Fermi}/LAT source catalogue (opened symbols). The down-triangles at the optical and $\gamma$-ray bands indicate the upper limits. The SEDs of knots are reproduced with the single-zone synchrotron+SSC+IC/CMB model, but the IC bumps are totally dominated by the IC/CMB components. In the last panel, the dashed line is derived by combining the model prediction fluxes of the 8 knots.}\label{LSJ}
\end{figure*}

The slopes of electron distribution, i.e., $p_1$ and $p_2$, can be constrained with the observed spectral indices at the radio and optical bands for knot-C, knot-E, and knot-IJ. For the other five knots, $p_1$ is also derived with their radio spectral indices, but their $p_2$ values are taken as $p_2=3.4$, which is derived by the average spectral index of the $\gamma$-ray emission from the core region in different states as shown in Figure \ref{core}. The values of $\gamma_{\rm min}$ can be constrained by comparing the predicted flux of the IC/CMB process with the observation data. A too small $\gamma_{\rm min}$ value would result in the predicted flux of the IC/CMB model in low-energy band exceeding the observations while a too large $\gamma_{\rm min}$ value could not explain the X-ray observation data. $\gamma_{\rm max}$ is fixed at a large value. Similar to the SED fitting of the core region, we also use the $\chi^{2}$ minimization technique to search the best SED fits. Our results are shown in Figure \ref{LSJ} and the derived model parameters with 1$\sigma$ confidence level are reported in Table 1.

As shown in Figure \ref{LSJ}, the $\gamma$-ray flux of each knot predicted by the IC/CMB process is lower than the observed steady $\gamma$-ray emission component. It indicates that a single knot may be insufficient to produce this emission component. Note that the low spatial resolution of \emph{Fermi}/LAT makes difficulty for judging the location of the $\gamma$-ray emission. Therefore, we add our SED fits of the eight knots to make a synthetical SED. It is also shown in Figure \ref{LSJ} in comparison with the spectra of the steady $\gamma$-ray emission component. It is found that the $\gamma$-ray flux of the synthetical SED is still lower than the observations of \emph{Fermi}/LAT during the steady $\gamma$-ray emission (MJD 55978--58142) with an integral flux ratio of 0.21. However, the $\gamma$-ray fluxes of the synthetical SED at several GeV energy band are roughly consistent with the observations. The steady emission component may be contributed by both the radio-core and large-scale knots though the contribution from knots is smaller than that of core region. Thus the fluctuations of the core emission would be concealed if the flux levels of emission from the radio-core and knots are comparable.

\begin{figure*}
\includegraphics[angle=0,scale=0.45]{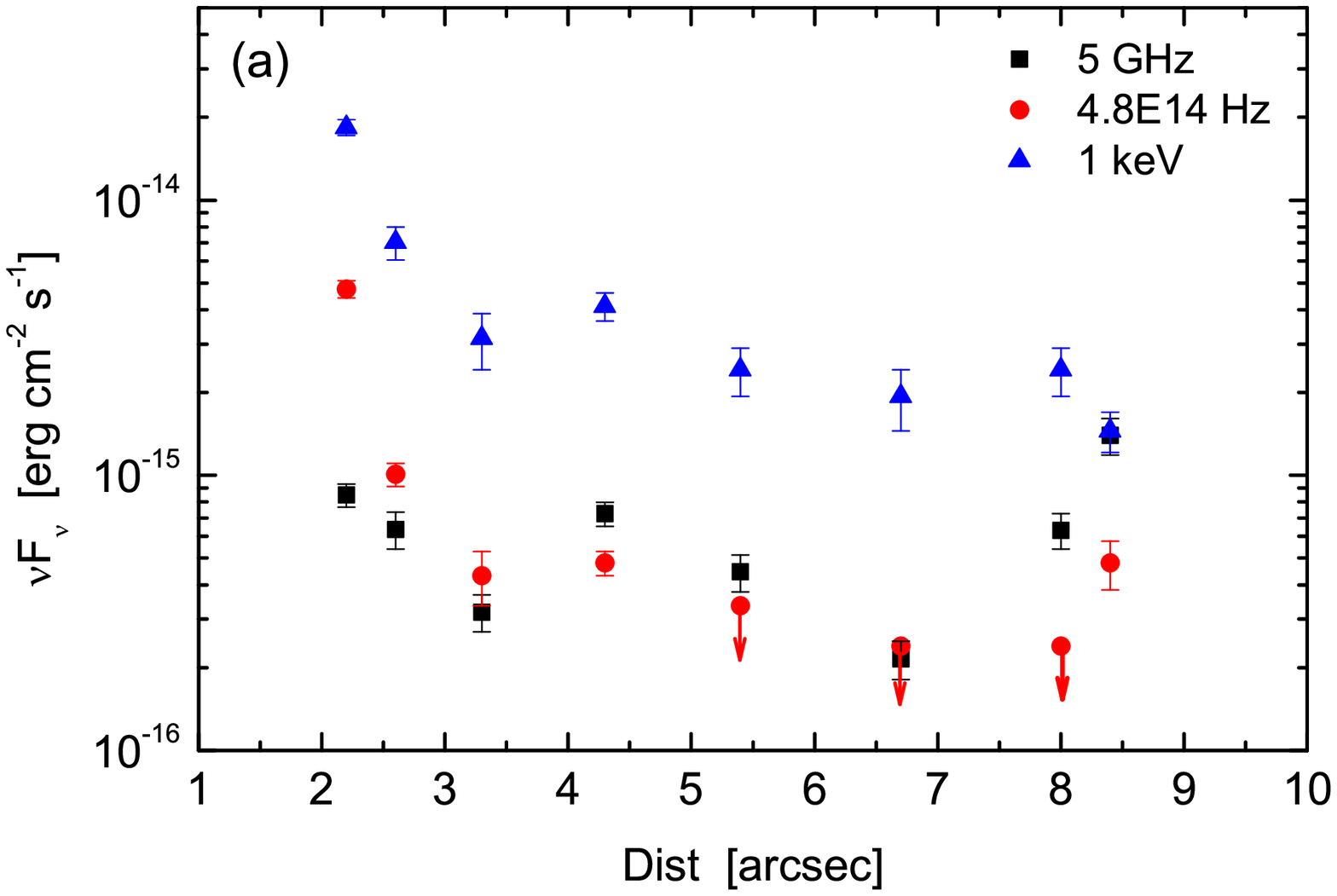}
\includegraphics[angle=0,scale=0.45]{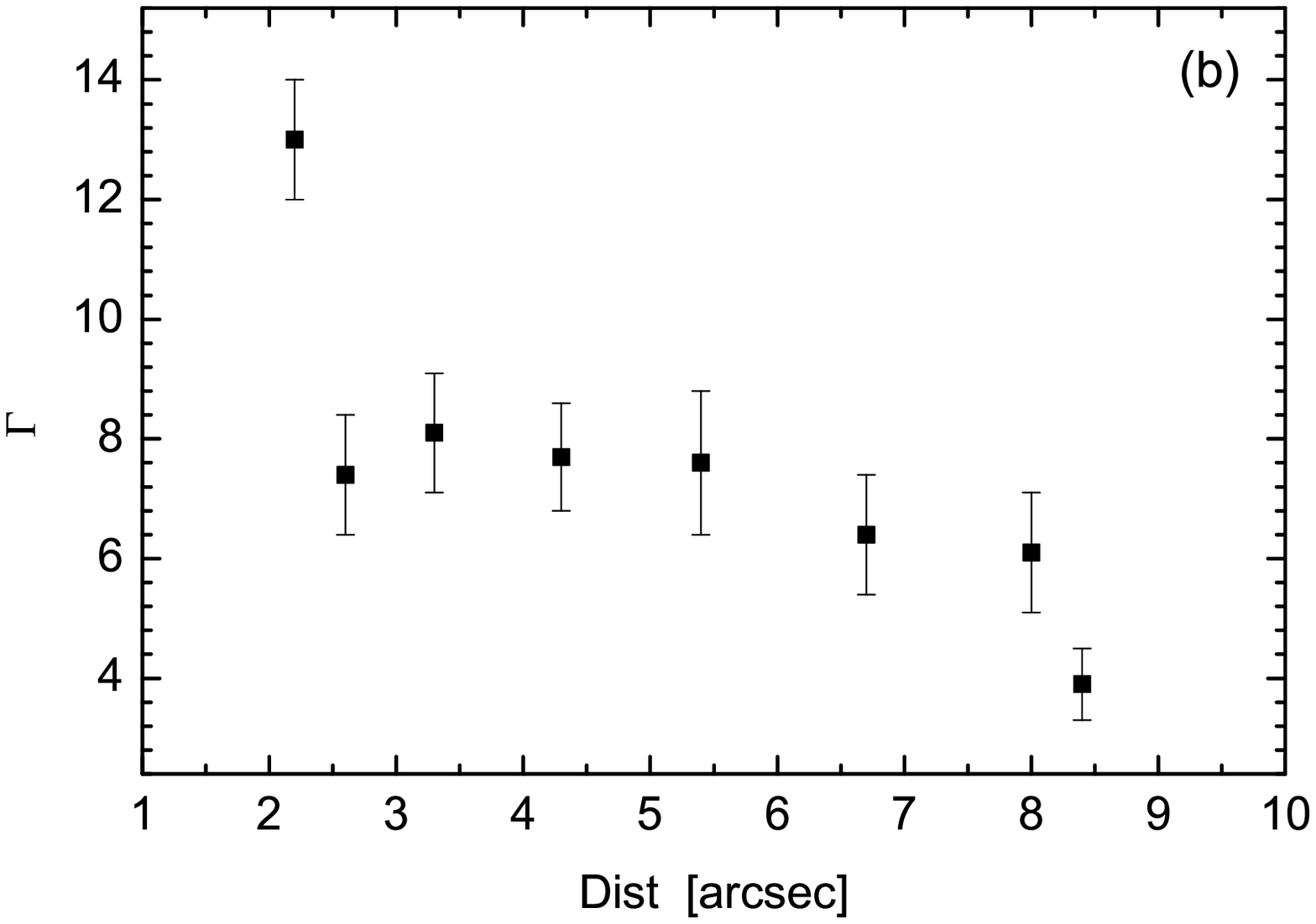}\\
\includegraphics[angle=0,scale=0.45]{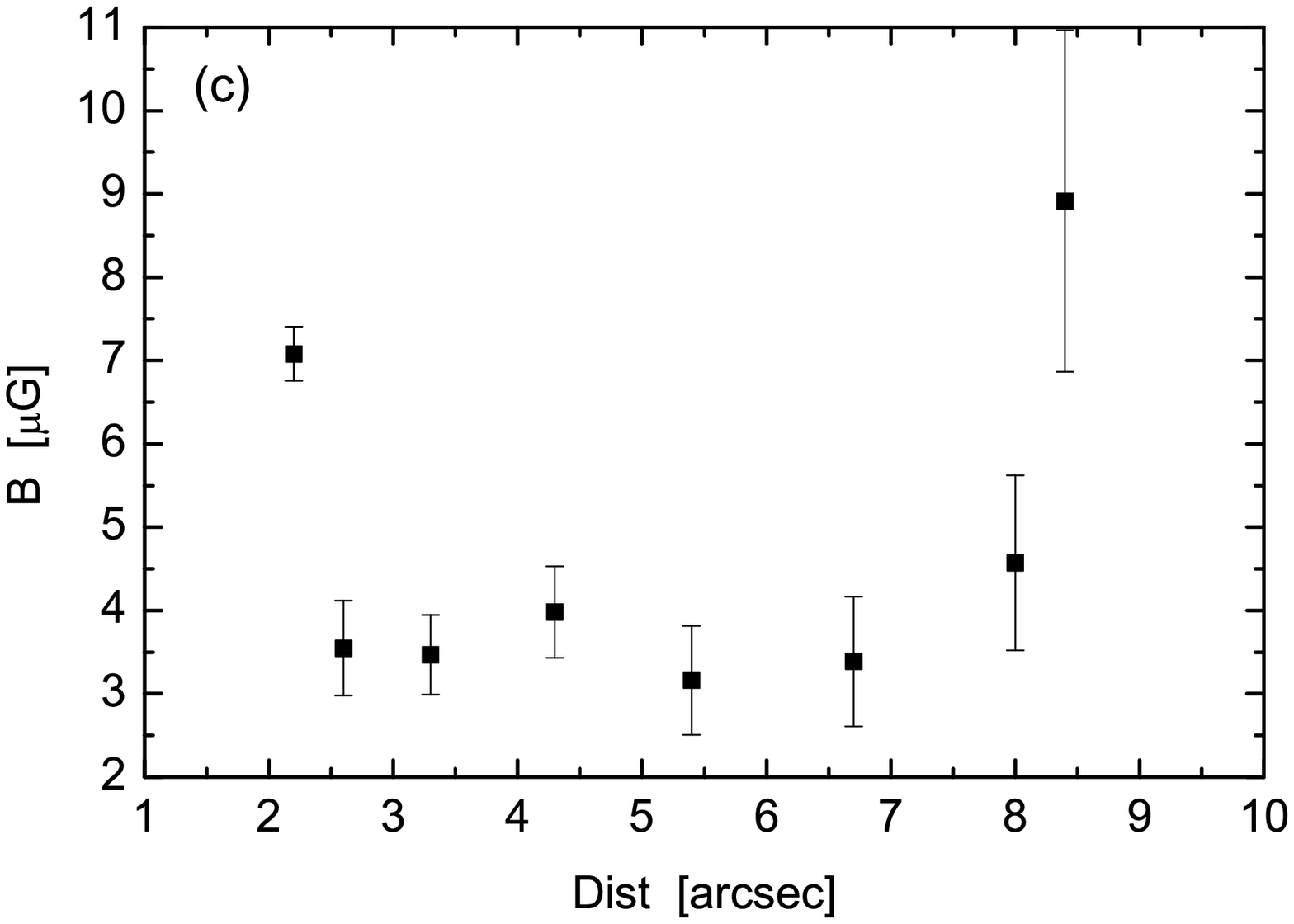}
\includegraphics[angle=0,scale=0.45]{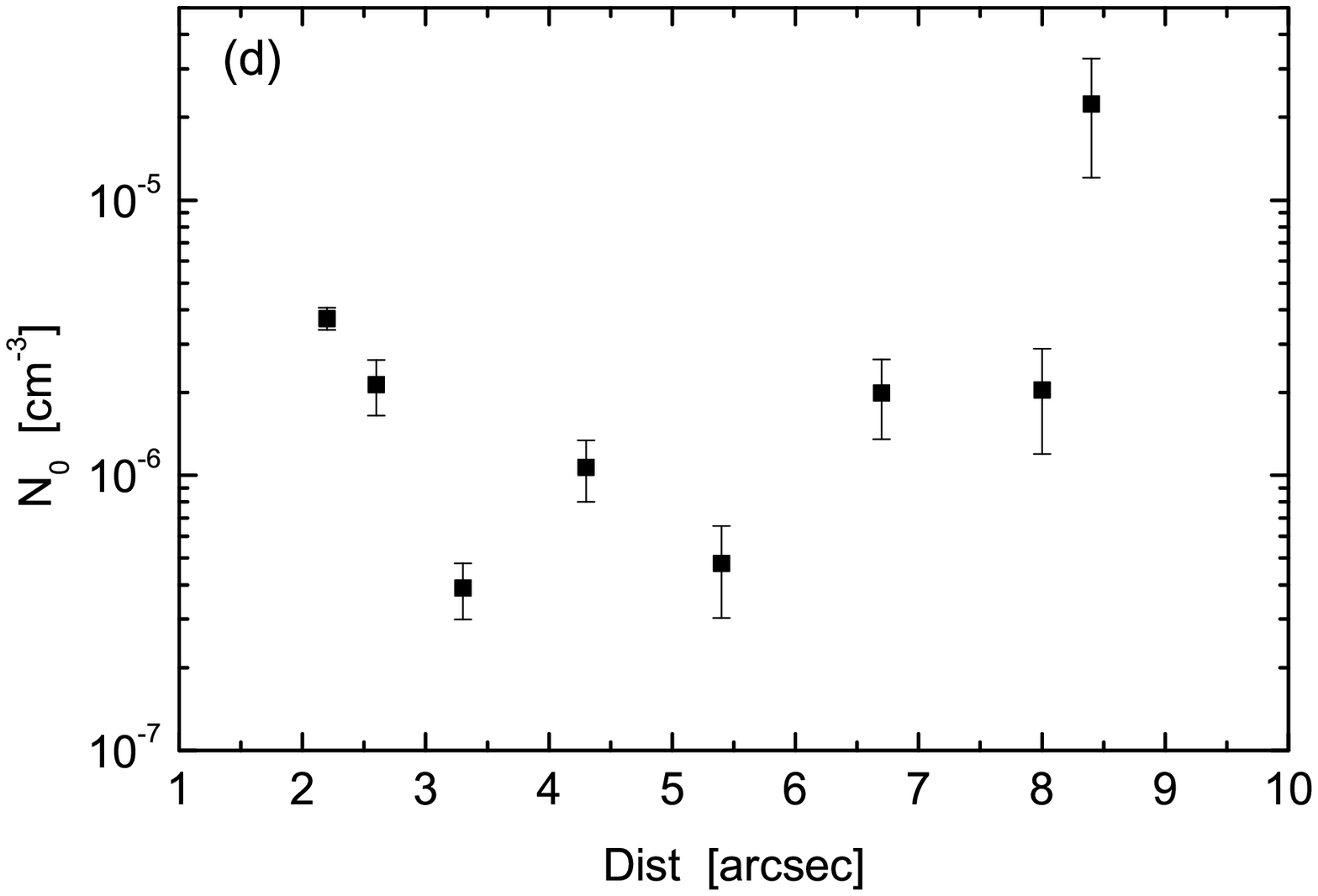}\\
\caption{Plots of the fluxes at 4.9 GHz, $4.8\times10^{14}$ Hz, and 1 keV, bulk Lorentz factor ($\Gamma$), magnetic field strength ($B$), and electron density parameter ($N_{0}$) along the jet of 4C +49.22. The X-axes indicate the distance from the core. }\label{dist}
\end{figure*}

The fluxes at 4.9 GHz, $4.8\times10^{14}$ Hz, 1 keV of these knots, together with the derived values of $\Gamma$, $B$, and $N_{0}$ by the model, against their distances from the radio-core are shown in Figure \ref{dist}. Except for the outmost knot-IJ, the fluxes at optical and X-ray bands together with the values of $\Gamma$ tend to decrease along the jet, but no similar trend is found for the radio flux, $B$, and $N_{0}$. This may be due to that the optical and X-ray radiations are produced by the IC/CMB process and the IC/CMB emission component is proportional to $\Gamma^2$, which makes $\Gamma$ a more efficient probe to the IC/CMB process than the other parameters. As reported in Sambruna et al. (2006), the X-ray jet of 4C +49.22 has a ``twisted" morphology, with a change in position angle of $\Delta$P.A. $\sim20^{\circ}$, which closely follows the radio morphology (Owen \& Puschell 1984). The plasma from the core region may be decelerated after the region that corresponds to knot-B due to the medium environments, and result in the ``twisted" jet morphology, which are also consistent with our analysis results.

As illustrated in Figure \ref{dist}, the outmost knot-IJ does not follow the same tendency as other knots; it has the highest $B$ and $N_0$ and the smallest $\Gamma$ among the eight knots. These features make knot-IJ more like a hotspot than a knot (see also Sambruna et al. 2006). As suggested by Zhang et al. (2010, 2018), the observed luminosity ratio of radio to X-ray, i.e., $L_{\rm 5~GHz}/L_{\rm 1~keV}$, is a characteristic to distinguish hotspots and knots. In the $L_{\rm 5~GHz}-L_{\rm 1~keV}$ plane, knot-IJ almost locates at the division line of $L_{\rm 1~keV}=L_{\rm 5~GHz}$. Therefore, at the jet terminal of 4C +49.22, i.e., knot-IJ, there is a significant increase of magnetic field and electron density and a decrease of bulk speed due to the interaction of the jet with circum medium.

The derived $\gamma_{\rm b}$ values of electrons in the knots are $\sim2\times10^4$--$3\times10^5$, corresponding to the typical energy of these electrons of $E_{\gamma_{\rm b}}\sim(10-154)$ GeV. The maximum energy is $E_{\gamma_{\rm max}}\sim(0.26-10.2)$ TeV, indicating that the electrons are effectively accelerated. The cooling time of electrons can be estimated with $t_{\rm cool}=3m_{\rm e}c^2/4\sigma_{\rm T}c\gamma_{e}(U^{'}_B+U^{'}_{\rm CMB}+U^{'}_{\rm syn})$, where $m_{\rm e}$ is the electron mass, $\sigma_{\rm T}$ is the Thomson cross section, $U^{'}_{\rm syn}$ is the synchrotron radiation energy density. $B$ ranges from 3.2 $\mu$G to 8.9 $\mu$G among the knots, and thus the corresponding $U^{'}_B$ is $\sim0.4-3.2\times10^{-12}$ erg cm$^{-3}$. In the comoving frame, $U^{'}_{\rm CMB}$ is much higher than $U^{'}_B$ and $U^{'}_{\rm syn}$. The IC/CMB process should dominate the cooling of electrons. We then obtain the light travel distance of $c\times t_{\rm cool}\simeq0.15-9.0$ kpc, which are shorter than the distance from the radio-core and even smaller than the size of knots. Therefore, the particles should be accelerated locally (see also in 3C 120; Zargaryan et al. 2017).

\section{Evolution of Jet Properties from Core to Knots}

On the basis of the model parameters derived from our SED fits, we estimate the power of each jet component by assuming that the jet consists of electrons, cold protons with one--to--one ratio, magnetic fields, and radiations. The total jet powers are carried by each component, i.e., $P_{\rm jet}=\pi R^2 \Gamma^2c(U^{'}_{\rm e}+U^{'}_{\rm p}+U^{'}_B+U^{'}_{\rm r})$. The energy density of radiation is calculated by $U^{'}_{\rm r}=L_{\rm obs}/4\pi R^2c\delta^4$. The derived jet powers together with the power carried by each component are reported in Table 2. The magnetized parameter ($\sigma_{B}$) and radiation efficiency ($\varepsilon_{\rm r}$) of jet are estimated by $\sigma_{B}=P_{B}/(P_{\rm r}+P_{\rm p}+P_{\rm e})$ and $\varepsilon_{\rm r}=P_{\rm r}/P_{\rm jet}$, and the derived values of $\sigma_{B}$ and $\varepsilon_{\rm r}$ for the core region and these knots are also given in Table 2.

For the core region, the jet powers are from $(4.0\pm0.8)\times10^{45}$ erg s$^{-1}$ to $(3.4\pm1.5)\times10^{46}$ erg s$^{-1}$ by considering the IC/BLR process, which are lower than the Eddington luminosity of $L_{\rm Edd}=5.0\times10^{46}$ erg s$^{-1}$ for $M_{\rm BH}=4.0\times10^8M_{\odot}$ (Shields et al. 2003) and $L_{\rm Edd}=2.0\times10^{47}$ erg s$^{-1}$ for $M_{\rm BH}=1.6\times10^9M_{\odot}$ (Decarli et al. 2008). The magnetized parameters of jet change from $0.012\pm0.007$ into $1.03\pm0.32$ while the radiation efficiencies of jet change from $0.006\pm0.005$ into $0.25\pm0.06$ following the increase of the $\gamma$-ray flux. Therefore, the core jet is highly magnetized with the high radiation efficiency at the high $\gamma$-ray emission state. However, the jet changes into the particle dominated at the low $\gamma$-ray emission state. As described in Section 3.1, the $\gamma$-ray radiation region may be away from the central engine following the decrease of the $\gamma$-ray flux, and the Poynting flux dominated jet changes into the particle dominated jet at the same time.

The estimated powers of the knots are in the range of $2.0\times10^{45}-1.4\times10^{48}$ erg s$^{-1}$. Calculations of the jet powers are sensitive to $\gamma_{\rm min}$ and most power is carried out by the protons for the small $\gamma_{\rm min}$ value under the assumption of the electron-proton number ratio as one--to--one. Note that the $\gamma_{\rm min}$ values of knots F, G, and H are constrained in a broad range being due to their poor observational data. Therefore, we give a range of the proton powers for the three knots in Table 2. The powers of knot-B and knot-D are $1.4\times10^{48}$ erg s$^{-1}$ and $8.2\times10^{46}$ erg s$^{-1}$, being exceeded the Eddington luminosity. This may be due to their $\gamma_{\rm min}$ values close to 1, and thus result in the so-called ``super-Eddington" jet powers as suggested by some authors (e.g., Dermer \& Atoyan 2004; Uchiyama et al. 2006; Meyer et al. 2015). As given in Table 2, both the magnetized parameters and radiation efficiencies are low for these knots, especially the radiation efficiencies that are much lower than that of the core region, indicating that these knots are particle dominant with very low radiation efficiencies.

\begin{figure*}
\includegraphics[angle=0,scale=0.45]{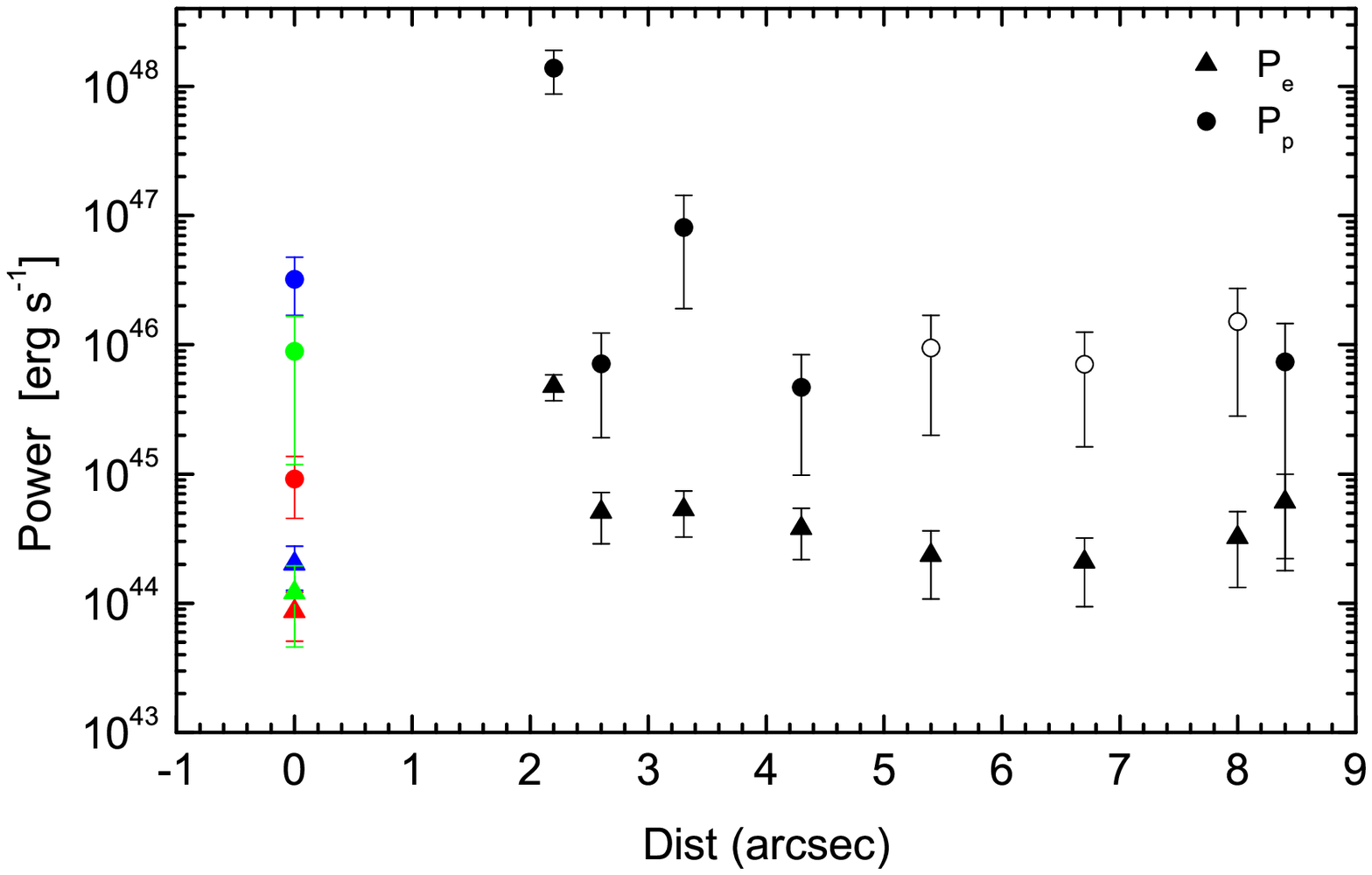}
\includegraphics[angle=0,scale=0.45]{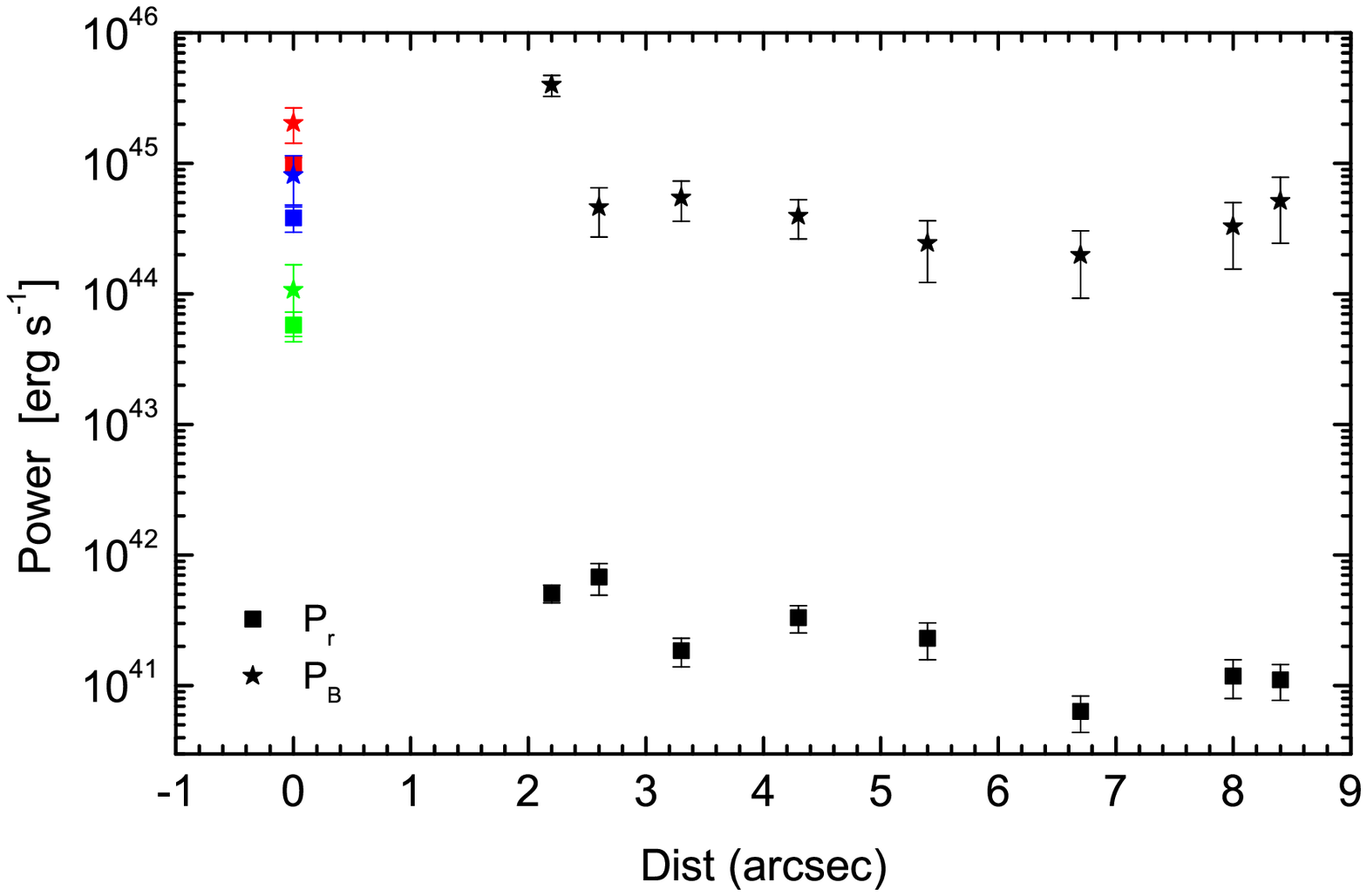}
\caption{Plot of the jet power carried by each component along the jet of 4C +49.22, triangles for $P_{\rm e}$, circles for $P_{\rm p}$, stars for $P_{B}$, and squares for $P_{\rm r}$. The X-axis indicates the distance from the core. The red, blue and green symbols are for the radio-core region at different epochs, i.e., at flare, post-flare, and low states, respectively. The black symbols are for the eight knots. The opened symbols in the left panel indicate the ranges of $P_{\rm p}$ for the three knots as listed in Table 2. }\label{power}
\end{figure*}

It is found that except for knot-B, the jet powers independently estimated for pc- and kpc-scale jets of 4C +49.22 are roughly in the same order, and the similar results had been reported for the $\gamma$-ray emission radio galaxy 3C 120 (Zargaryan et al. 2017). We plot the derived power carried by each component against the distance from the radio-core in Figure \ref{power}. Comparing with the core region, the jet powers of knots are dominated by the particle powers. As suggested in Zargaryan et al. (2017), the jet may not substantially dissipate its power until its end, but it becomes radiatively inefficient in large-scale comparing with the core region. In addition, the Poyting flux dominating jet in pc-scale convert to the particle dominating jet in kpc-scale through some unknown mechanisms, and the jet decelerates in large-scale by interacting with surrounding medium (Boccardi et al. 2017; Chen 2018).

\section{Discussion }

\begin{figure*}
\includegraphics[angle=0,scale=1.8]{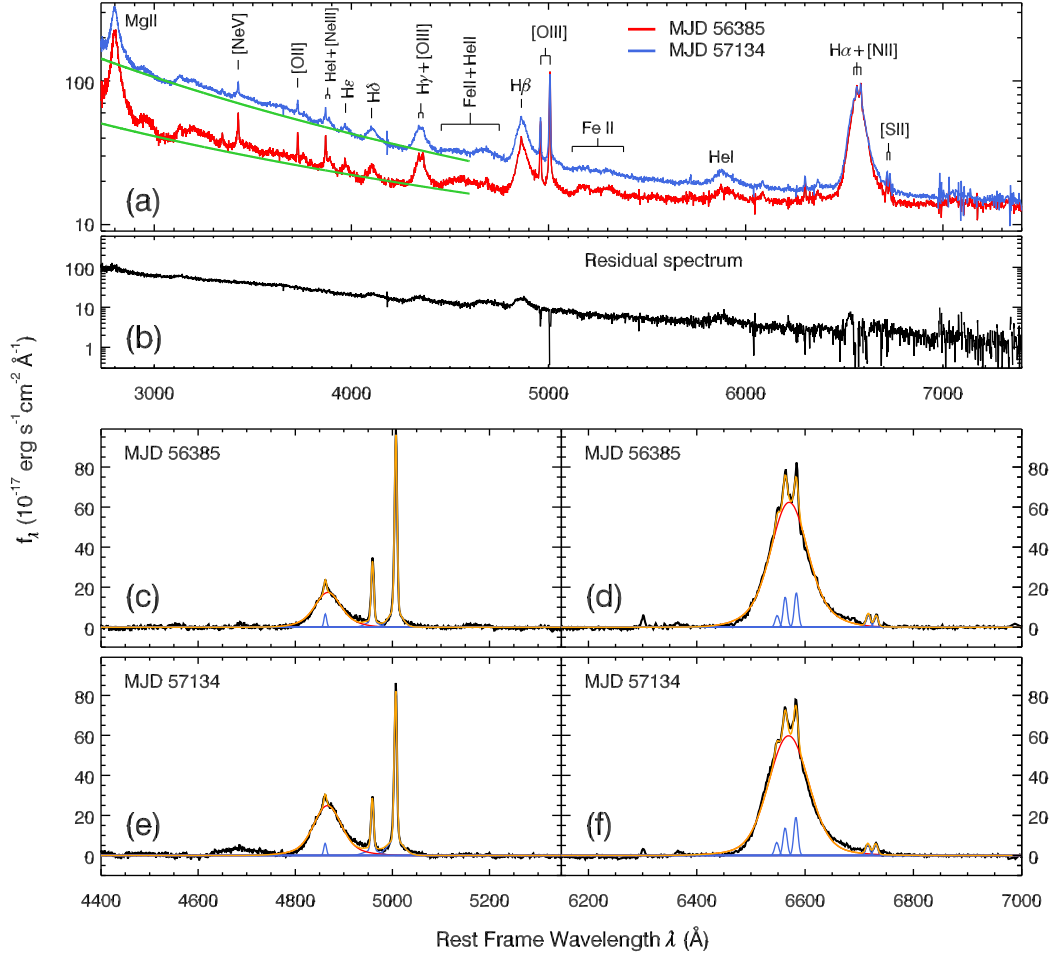}
\caption{Panel (a): SDSS spectra of 4C +49.22 at epochs of MJD 56385 and MJD 57134, respectively, with the visible emission lines labeled. The green lines represent the best-fitting power-law continuum to each spectrum at $<4200$~{\AA}. Panel (b): Residuals by subtracting the spectrum at MJD 56385 from the spectrum at MJD 57134. Panels (c)--(f): Residuals (in black) of the spectrum in H$\beta$+O{\scriptsize III} region (left) and H$\alpha$+N{\scriptsize II}+S{\scriptsize II} region (right) after subtracting the power-law continuum and the Fe{~\scriptsize II} model. The best-fitting models of the emission lines are displayed in red and blue for the broad and narrow components, respectively, and their sum is in orange.   }\label{SDSS}
\end{figure*}

The multiwavelength observations are useful to judge the origin of the $\gamma$-ray emission. 4C +49.22 was spectroscopically observed two times in the Sloan Digital Sky Survey (SDSS) during the \emph{Fermi}/LAT observations, however, both of them were obtained when the source is in the steady $\gamma$-ray state, i.e., at MJD 56385 and MJD 57134, respectively. After correcting for the Galactic extinction and transforming into the source rest-frame, we analyze its SDSS spectra following the same approach adopted in Yao et al. (2015)\footnote{The Balmer lines consist of a broad and a narrow components. The broad component is modeled by the double Gaussian profiles while the narrow component is modeled by a single Gaussian profile. Two Gaussian components are used to model each line of the O{\sc\,iii}~$\lambda$4959, $\lambda$5007 doublet, one for the center and one for the blue-shifted wing component. The N{\sc\,ii}~$\lambda$6548, $\lambda$6583 doublet and the S{\sc\,ii}~$\lambda$6716, $\lambda$6731 doublet are modeled by a single Gaussian profile. The flux ratios of both O{\sc\,iii}~$\lambda$4959, $\lambda$5007 and N{\sc\,ii}~$\lambda$6548, $\lambda$6583 are fixed at the theoretical values.}. A broken power-law is used to model the continuum emission and the Fe{\sc\,ii} multiplets are modeled with the templates in V\'{e}ron-Cetty et al. (2004). The results of spectroscopical analysis in SDSS observations are given in Figure \ref{SDSS} and Table 3.

An obvious bluer--when--brighter trend can be found for the two spectroscopical observation epochs, however, the fluxes of the broad-lines keep constants when the continuum flux increases. The continuum flux is estimated by the flux at 3000~{\AA} and its ratio is $\sim$2.6 for the two epochs. With a radio-quiet AGN sample, Ai et al. (2010) reported that the continuum variation from the accretion disk is weaker for the brighter sources. The derived disk luminosity by fitting the blue bump when 4C +49.22 was at a low state is 6.4$\times10^{45}$ erg s$^{-1}$, which likely indicates the smaller variability amplitude of the accretion disk. And no variation of broad-line flux also indicates that the continuum variation may be due to the beamed jet contribution, not the unbeamed thermal disk radiation. This is also consistent with the broadband SED fitting results of the core region; the disk radiation component of 4C +49.22 keeps constant at different $\gamma$-ray emission states.

As displayed in Figure \ref{LC}, the $\gamma$-ray fluxes are very low at MJD 56385 and MJD 57134. No counterpart in the $\gamma$-ray emission to the flare in the low-energy bands may be caused by a variation of the magnetic field in the emitting region (Chatterjee et al. 2013). On the other hand, it likely demonstrates that the steady $\gamma$-ray emission component of 4C +49.22 is partly contributed by the large-scale knots. Note that there is a He{\sc\,ii} $\lambda$4687 component in the spectrum at MJD 57134, which is not seen in the spectrum at MJD 56385. If the BLR is radial ionization stratification as reported by Poutanen \& Stern (2010), this implies that the radiation regions at different epochs may locate at different distances from the black hole (see also 3C 273 in Pati\~{n}o-\'{A}lvarez et al. 2018). We will not give more discussion on this issue since the derived He{\sc\,ii} $\lambda$4687 component is not significant.

\begin{figure*}
\includegraphics[angle=0,scale=0.45]{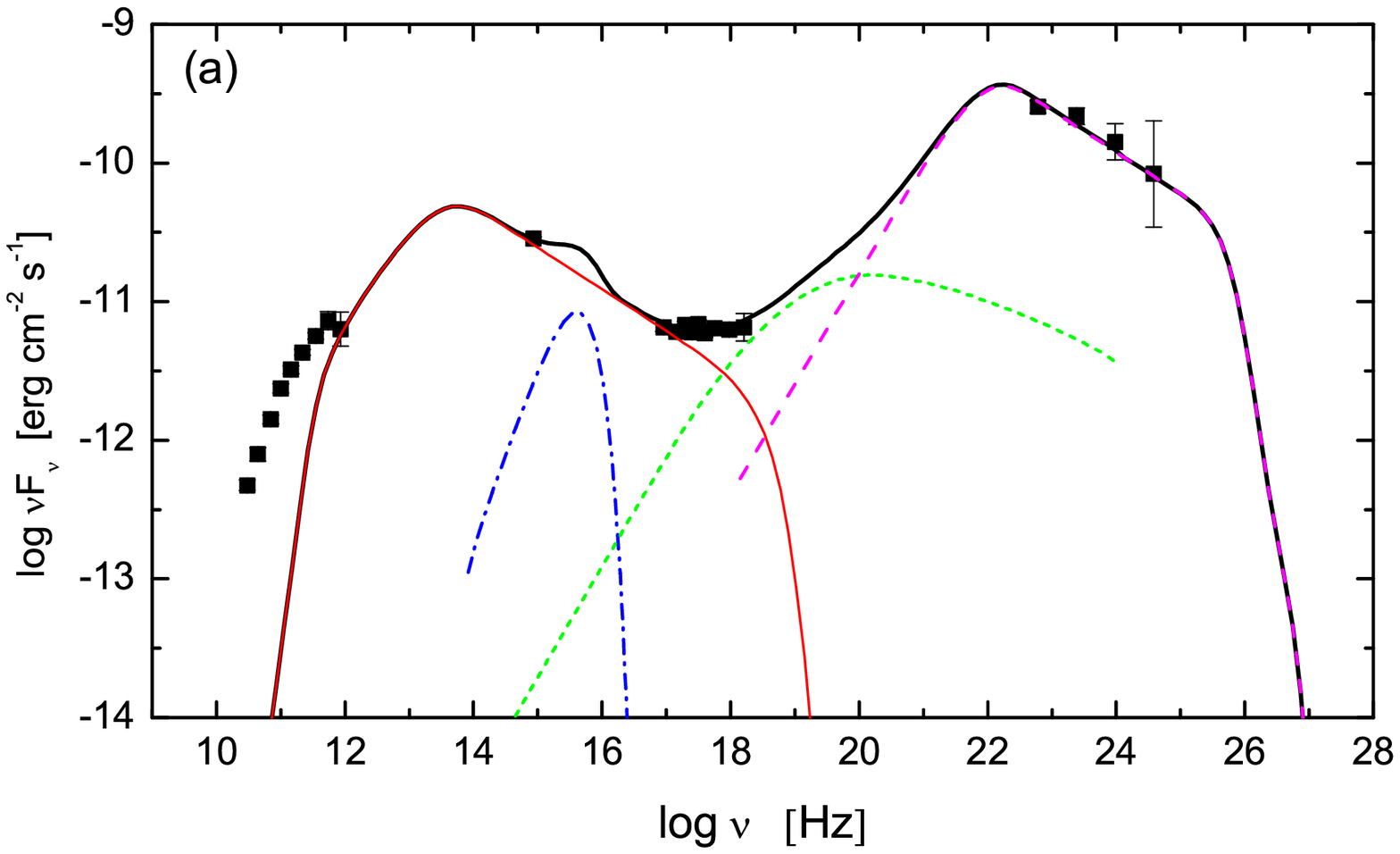}
\includegraphics[angle=0,scale=0.45]{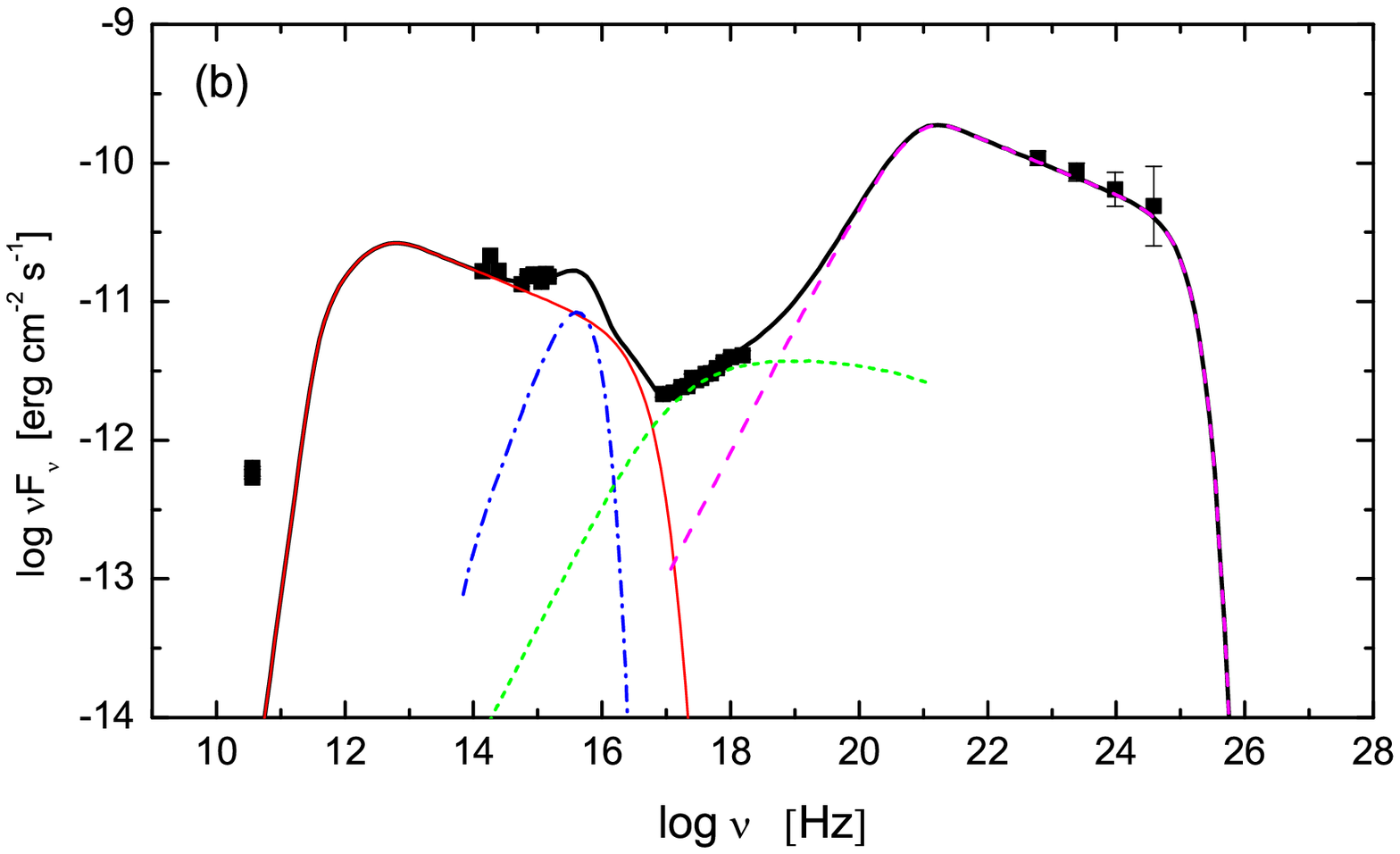}
\includegraphics[angle=0,scale=0.45]{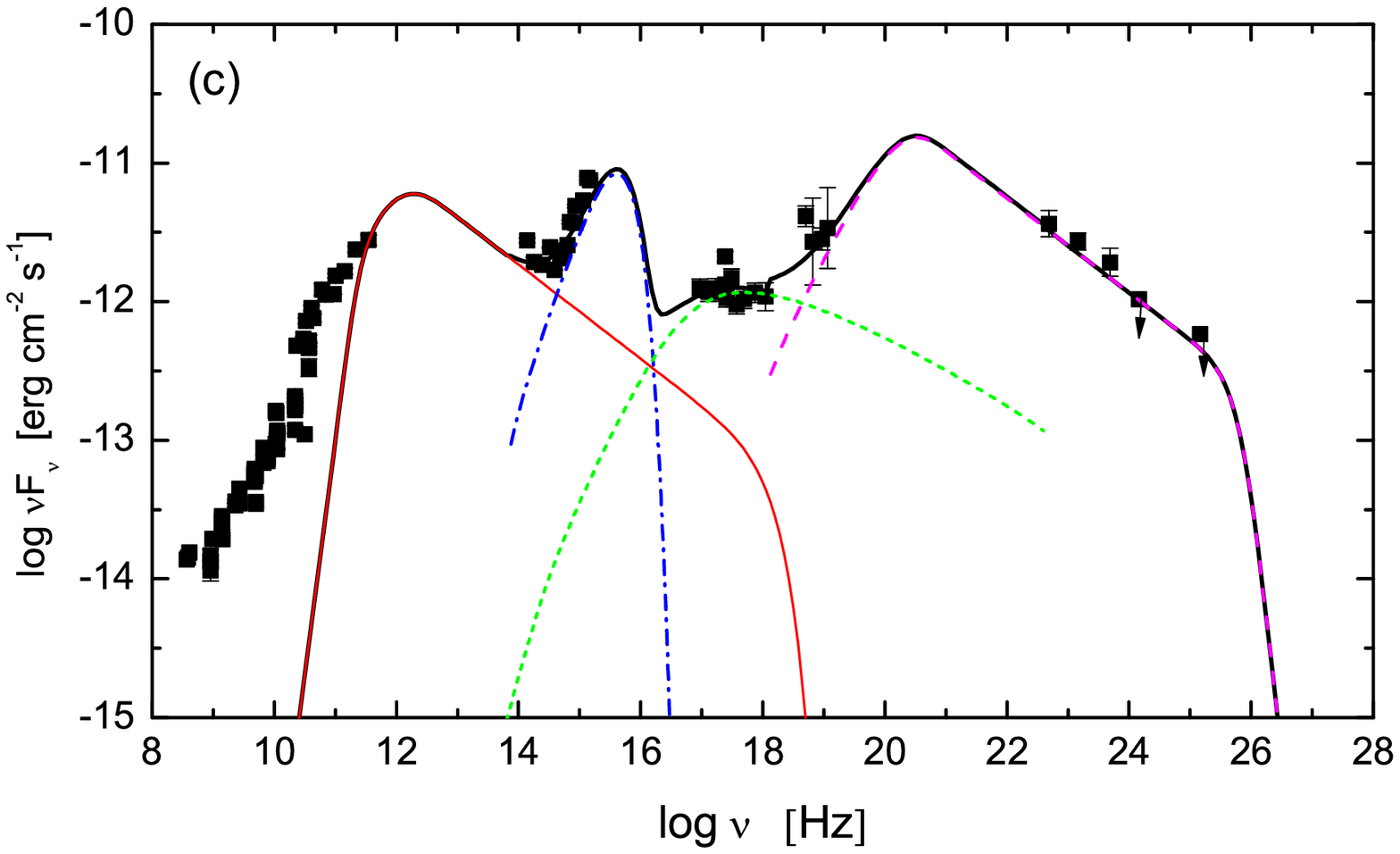}
\caption{The same symbols as in Figure \ref{core}, but the EC bumps are reproduced by the IC/torus process (magenta dashed lines), not the IC/BLR process. }\label{core_T}
\end{figure*}

Although the increase of flux density and polarization degree in the radio-core of 4C +49.22 after the $\gamma$-ray flare implies that the $\gamma$-ray emission is produced in or close the radio-core (Cutini et al. 2014), we cannot certainly judge that the $\gamma$-ray radiation region is within or without the BLR. Hence, we also consider the IC/torus process to explain the $\gamma$-ray emission by assuming that the $\gamma$-ray radiation region is outside of the BLR. In this scenario, the corresponding photon field energy density in the comoving frame is $U^{'}_{\rm IR}=3\times10^{-4}\Gamma^2$ erg cm$^{-3}$. The spectrum of torus is assumed to be a blackbody with a peak frequency in the comoving frame at $3\times10^{13}\Gamma$ Hz (Ghisellini \& Tavecchio 2008). Similar to the IC/BLR scenario, the assumption of $\delta=\Gamma$ is taken during the SED fitting for the $\gamma$-ray flare, and then a viewing angle of $\theta\sim2.9^{\circ}$ is obtained and fixed during SED modeling for post-flare and low states. The fitting results are given in Figure \ref{core_T} and the derived parameters and jet powers are also listed in Tables 1 and 2.

Comparing with the IC/BLR scenario, $\Gamma$ is larger and $B$ is smaller in the IC/torus case, and the radio-core jet is likely dominated by the particle powers with the low radiation efficiency. The single-zone synchrotron+SSC+IC/torus model can also represent the observed SEDs of the core region well. However, a harder electron spectrum, which cannot be explained with the first-order Fermi acceleration via relativistic shocks (e.g., Kirk et al. 2000; Achterberg et al. 2001; Virtanen \& Vainio 2005). In this respect, we suggest that the synchrotron+SSC+IC/BLR model is the preferred model to explain the SEDs of the core region.

Note that the photon fields of accretion disk, torus, and BLR were taken into account together to calculate the EC process in Cutini et al. (2014). We used the different electron distribution spectrum and radiation models to explain the SEDs of the core region, therefore, we obtained the different fitting parameters from them. The values of $B$ and $\delta$ reported in Cutini et al. (2014) are more close to the derived values in the IC/torus case of our work.

\section{Conclusions}

We dealt with and analyzed the long-term monitoring data of 4C +49.22 by \emph{Fermi}/LAT, besides a large outburst, and a long-term, steady $\gamma$-ray emission component is observed in the $\gamma$-ray light curve, which almost can be fitted by a constant flux. The broadband SEDs of the core region at different $\gamma$-ray emission epochs (during the large flare, post-flare, low state) can be well reproduced with the single-zone leptonic model, synchrotron+SSC+IC/BLR. The SEDs of the eight knots in large-scale can also be well explained with single-zone leptonic model, synchrotron+SSC+IC/CMB, and the $\gamma$-ray fluxes predicted by the model are lower than that of the steady $\gamma$-ray emission observed with \emph{Fermi}/LAT. The synthetical fluxes predicted by the model for the eight knots at the \emph{Fermi}/LAT energy band are still lower than this steady $\gamma$-ray emission with the integral flux ratio of 0.21. It indicates that the steady $\gamma$-ray emission is still dominated by the core radiation, but may be partially contributed by the large-scale knots, which may conceal the low-level flux variation of the $\gamma$-ray emission from the core region. The decreases of the flux at X-ray and optical bands and the derived bulk Lorentz factors along the jet in large-scale may indicate that the jet decelerates in large-scale by interacting with surrounding medium.

On the basis of the fitting parameters, we calculated the jet powers and the power carried by each component in pc-scale and kpc-scale. It was found that the jet powers independently estimated for pc- and kpc-scale jets of 4C +49.22 are roughly in the same order. The magnetization parameters and radiation efficiencies of the core region decrease with the decrease of the $\gamma$-ray emission flux, and the jet changes from Poynting flux dominated into particle dominated. Comparing with the core region, the knots in large-scale are particle dominated with very low radiation efficiencies. Therefore, the Poyting flux dominating jet in pc-scale converts to the particle dominating jet in kpc-scale through some unknown mechanisms. The synchrotron+SSC+IC/CMB model indeed induces the `super-Eddington" jet powers for some sources as suggested by some authors (e.g., Dermer \& Atoyan 2004; Uchiyama et al. 2006; Meyer et al. 2015).

\acknowledgments

We thank the anonymous referee for his/her valuable suggestions. We thank Dr. Sara Cutini for providing us the observation data of the core region and we also thank the helpful discussion with Dr. Yuan Liu. This work is supported by the National Natural Science Foundation of China (grants 11573034, 11533003, 11851304, and U1731239), the National Basic Research Program (973 Programme) of China (grant 2014CB845800). En-Wei Liang acknowledges support from the special funding from the Guangxi Science Foundation for Guangxi distinguished professors (grant 2017AD22006 for Bagui Yingcai \& Bagui Xuezhe). Su Yao acknowledges the support by the KIAA-CAS Fellowship, which is jointly supported by Peking University and Chinese Academy of Sciences.


\clearpage

\begin{deluxetable}{lcccccccccc}
\tabletypesize{\scriptsize} \rotate  \tablecolumns{11}\tablewidth{37pc} \tablecaption{SED Modelling Parameters for the Core Region and Knots}\tablenum{1} \tablehead{\colhead{Comp\tablenotemark{a}}&\colhead{\tiny{$\gamma_{\rm min}$}}&\colhead{\tiny{$\log\gamma_{\rm b}$}}&\colhead{\tiny{$\gamma_{\rm max}$}}&\colhead{\tiny{$\log N_{0}$}}&\colhead{$p_{1}$} &\colhead{$p_{2}$}& \colhead{$\Gamma$}&\colhead{$B$}&\colhead{$R$}\\
\colhead{}&\colhead{}&\colhead{}&\colhead{}&\colhead{[cm$^{-3}$]}&\colhead{}&\colhead{}
&\colhead{}& \colhead{\tiny{[G]}}&\colhead{[cm]}}
\startdata
F$^{\rm B}$&59$\pm$25&2.55$\pm$0.14&5E4&4.61$\pm$0.11&2&3.7&13.2$\pm$0.8&6.7$\pm$0.7&(8.46$\pm$0.51)E15\\
PF$^{\rm B}$&2$\pm$1&2.16$\pm$0.10&1E4&4.67$\pm$0.10&2&3.38&9.1$\pm$1.0&4.3$\pm$0.8&(1.21$\pm$0.05)E16\\
L$^{\rm B}$&6$\pm$4&2.29$\pm$0.11&5E4&5.81$\pm$0.12&2.4&3.8&3.9$\pm$0.5&3.3$\pm$0.7&(1.38$\pm$0.15)E16\\
F$^{\rm T}$&1$\pm$0&2.86$\pm$0.10&2E5&2.85$\pm$0.13&1.4&3.6&20.0$\pm$1.5&1.05$\pm$0.20&(1.28$\pm$0.09)E16\\
PF$^{\rm T}$&1$\pm$0&2.34$\pm$0.10&2.3E4&2.55$\pm$0.08&1.2&3.38&16.5$\pm$1.2&0.96$\pm$0.10&(1.89$\pm$0.02)E16\\
L$^{\rm T}$&12$\pm$6&2.35$\pm$0.12&2E5&2.77$\pm$0.14&1.2&3.68&7.0$\pm$0.8&0.59$\pm$0.18&(2.41$\pm$0.22)E16\\
Knot-B&2.5$\pm$0.5&4.35$\pm$0.30&5E5&-5.43$\pm$0.04&2.68&3.4\tablenotemark{b}&13.0$\pm$1.0&(7.1$\pm$0.3)E-6&1.12E22$^{*}$\\
Knot-C&40$\pm$20&5.44$\pm$0.11&2E7&-5.67$\pm$0.10&2.5&3.7&7.4$\pm$1.0&(3.5$\pm$0.6)E-6&1.34E22$^{*}$\\
Knot-D&3$\pm$2&4.50$\pm$0.50&6E5&-6.41$\pm$0.10&2.4&3.4\tablenotemark{b}&8.1$\pm$1.0&(3.5$\pm$0.5)E-6&1.34E22$^{*}$\\
Knot-E&35$\pm$25&4.45$\pm$0.13&5.6E6&-5.97$\pm$0.11&2.36&3.3&7.7$\pm$0.9&(4.0$\pm$0.5)E-6&1.05E22$^{*}$\\
Knot-F&10--60&4.38$\pm$0.35&5E5&-6.32$\pm$0.16&2.3&3.4\tablenotemark{b}&7.6$\pm$1.2&(3.2$\pm$0.7)E-6&1.05E22$^{*}$\\
Knot-G&15--70&4.35$\pm$0.35&5E5&-5.70$\pm$0.14&2.5&3.4\tablenotemark{b}&6.4$\pm$1.0&(3.4$\pm$0.8)E-6&1.05E22$^{*}$\\
Knot-H&10--60&4.49$\pm$0.37&5E5&-5.69$\pm$0.18&2.42&3.4\tablenotemark{b}&6.1$\pm$1.0&(4.6$\pm$1.1)E-6&1.05E22$^{*}$\\
Knot-IJ&45$\pm$30&5.28$\pm$0.11&4.2E6&-4.65$\pm$0.20&2.56&4.36&3.9$\pm$0.6&(8.9$\pm$2.0)E-6&1.05E22$^{*}$\\
\enddata
\tablecomments{The superscript $^{*}$ in column of $R$ indicates that it is frozen.}
\tablenotetext{a}{``F", ``PF", and ``L" indicating the flare, post-flare, and the archival low-state of the $\gamma$-ray emission, respectively. The superscripts denote the different IC processes, i.e., ``B" for IC/BLR and ``T" for IC/torus, respectively. }
\tablenotetext{b}{It is fixed and derived by the average spectral index of the $\gamma$-ray emission from the core region in different states as shown in Figure \ref{core}. }
\end{deluxetable}

\begin{deluxetable}{lccccccc}
\tabletypesize{\scriptsize} \rotate  \tablecolumns{8}\tablewidth{40pc} \tablecaption{Derived Jet Power and Power Carried by Each Component}\tablenum{2} \tablehead{\colhead{Comp\tablenotemark{a}}&\colhead{\tiny{$\log P_{\rm e}$}}&\colhead{\tiny{$\log P_{\rm p}$}}&\colhead{\tiny{$\log P_{B}$}}&\colhead{\tiny{$\log P_{\rm r}$}}&\colhead{\tiny{$\log P_{\rm jet}$}} &\colhead{$\varepsilon_{\rm r}$}& \colhead{$\sigma_{B}$}\\
\colhead{}&\colhead{\tiny{[erg s$^{-1}$]}}&\colhead{\tiny{[erg s$^{-1}$]}}&\colhead{\tiny{[erg s$^{-1}$]}}&\colhead{\tiny{[erg s$^{-1}$]}}&\colhead{\tiny{[erg s$^{-1}$]}}&\colhead{}
&\colhead{}}
\startdata
F$^{\rm B}$&43.94$\pm$0.18&44.96$\pm$0.22&45.31$\pm$0.13&44.99$\pm$0.05&45.60$\pm$0.08&0.25$\pm$0.06&1.03$\pm$0.32\\
PF$^{\rm B}$&44.30$\pm$0.16&46.51$\pm$0.21&44.91$\pm$0.18&44.58$\pm$0.10&46.53$\pm$0.20&0.011$\pm$0.006&0.025$\pm$0.010\\
L$^{\rm B}$&44.08$\pm$0.27&45.95$\pm$0.38&44.03$\pm$0.24&43.76$\pm$0.11&45.96$\pm$0.37&0.006$\pm$0.005&0.012$\pm$0.007\\
F$^{\rm T}$&44.59$\pm$0.19&46.16$\pm$0.18&44.40$\pm$0.20&44.76$\pm$0.07&46.20$\pm$0.17&0.036$\pm$0.015&0.016$\pm$0.007\\
PF$^{\rm T}$&44.56$\pm$0.14&46.21$\pm$0.11&44.51$\pm$0.12&44.40$\pm$0.06&46.24$\pm$0.10&0.015$\pm$0.004&0.019$\pm$0.005\\
L$^{\rm T}$&44.15$\pm$0.26&45.48$\pm$0.24&43.49$\pm$0.33&43.42$\pm$0.10&45.51$\pm$0.23&0.008$\pm$0.005&0.010$\pm$0.007\\
Knot-B&45.68$\pm$0.10&48.14$\pm$0.16&45.60$\pm$0.08&41.71$\pm$0.07&48.15$\pm$0.16&(3.6$\pm$1.5)E-7&(2.9$\pm$0.5)E-3\\
Knot-C&44.70$\pm$0.19&45.85$\pm$0.32&44.66$\pm$0.18&41.83$\pm$0.12&45.91$\pm$0.28&(8.4$\pm$5.9)E-5&0.06$\pm$0.02\\
Knot-D&44.73$\pm$0.17&46.91$\pm$0.33&44.74$\pm$0.15&41.27$\pm$0.11&46.91$\pm$0.33&(2.3$\pm$1.8)E-6&0.007$\pm$0.002\\
Knot-E&44.58$\pm$0.19&45.67$\pm$0.34&44.60$\pm$0.15&41.52$\pm$0.10&45.74$\pm$0.30&(6.0$\pm$4.3)E-5&0.08$\pm$0.03\\
Knot-F\tablenotemark{b}&44.37$\pm$0.24&45.30--46.23&44.39$\pm$0.22&41.36$\pm$0.14&45.39--46.24&(1.3--9.3)E-5&0.01--0.11\\
Knot-G\tablenotemark{b}&44.32$\pm$0.24&45.21--46.10&44.30$\pm$0.23&40.80$\pm$0.14&45.31--46.11&(0.5--3.1)E-5&0.02--0.11\\
Knot-H\tablenotemark{b}&44.51$\pm$0.26&45.45--46.44&44.51$\pm$0.23&41.08$\pm$0.14&45.54--46.45&(0.4--3.4)E-5&0.01--0.10\\
Knot-IJ&44.78$\pm$0.28&45.87$\pm$0.42&44.71$\pm$0.23&41.05$\pm$0.13&45.93$\pm$0.37&(1.3$\pm$1.2)E-5&0.06$\pm$0.03\\
\enddata
\tablenotetext{a}{``F", ``PF", and ``L" indicating the flare, post-flare, and the archival low-state of the $\gamma$-ray emission, respectively. The superscripts denote the different IC processes, i.e., ``B" for IC/BLR and ``T" for IC/torus, respectively. }
\tablenotetext{b}{Their powers of protons are derived by the two sets of parameters with the smallest and largest $\gamma_{\rm min}$ values. Hence, the ranges of $P_{\rm p}$, $P_{\rm jet}$, $\varepsilon_{\rm r}$, and $\sigma_{B}$ are presented for the three knots. }
\end{deluxetable}
\begin{deluxetable}{lcccccccccccc}
\tabletypesize{\scriptsize} \rotate  \tablecolumns{14}\tablewidth{44pc} \tablecaption{SDSS Observations}\tablenum{3} \tablehead{\colhead{Name\tablenotemark{1}}& \multicolumn{2}{c}{H$\alpha~\lambda6563$$^{\rm B}$}&\multicolumn{2}{c}{H$\alpha~\lambda6563$$^{\rm N}$}&\multicolumn{2}{c}{H$\beta~\lambda4861$$^{\rm B}$}&\multicolumn{2}{c}{H$\beta~\lambda4861^{\rm N}$}&\multicolumn{2}{c}{O{\sc\,iii}$~\lambda5007$
}&\multicolumn{2}{c}{$\lambda$3000~{\AA}}}

\startdata
Time\tablenotemark{2}&56385&57134&56385&57134&56385&57134&56385&57134&56385&57134&56385&57134\\
FWHM\tablenotemark{3}&3763.2&4011.8&383.7&383.7&3982.5&4226.3&376.7&391.7&396.3&410.8&\nodata&\nodata\\
Flux\tablenotemark{4}&7.83&7.83&0.19&0.18&1.73&2.63&0.06&0.06&1.18&1.12&1.65&4.26\\
EW\tablenotemark{5}&413.4&348.6&9.8&7.8&80.5&74.5&2.8&1.6&56&33.5&\nodata&\nodata\\

\enddata
\tablenotetext{1}{The superscripts `B' and `N' indicate the broad and narrow components of emission lines.}
\tablenotetext{2}{The observation time in units of MJD.}
\tablenotetext{3}{Full width at half maximum in units of km~s$^{-1}$. }
\tablenotetext{4}{The flux in units of 10$^{-14}$~erg~s$^{-1}$~cm$^{-2}$ for the emission lines and 10$^{-12}$~erg~s$^{-1}$~cm$^{-2}$ for the continuum at 3000~{\AA}, respectively.}
\tablenotetext{5}{Equivalent width in units of {\AA}.}
\end{deluxetable}

\end{document}